\documentclass[12pt]{article}
\usepackage{a4wide}
\usepackage{amssymb,amsmath,bm}
\usepackage{setspace}
\usepackage{natbib}
\usepackage{graphicx}
\usepackage{subcaption}
\usepackage{color}
\usepackage{mdframed}
\usepackage{amsthm}
\usepackage{fancyvrb}
\usepackage{multirow,bigdelim,siunitx}
\usepackage{hyperref}
\usepackage{listings}
\usepackage{algpseudocode}
\usepackage{algorithm}
\usepackage{booktabs}

\algtext*{EndFor}

\newtheorem{result}{Result}

\makeatletter
\newcommand*{\addFileDependency}[1]{
  \typeout{(#1)}
  \@addtofilelist{#1}
  \IfFileExists{#1}{}{\typeout{No file #1.}}
}
\makeatother

\newcolumntype{d}[1]{D{.}{.}{#1}}

\newcommand{\chref}[2]{\href{#1}{\textcolor{blue}{#2}}}

\title{Multivariate MRP}
\author{Max Goplerud and Michael Auslen\thanks{We thank Jim Bisbee, Jacob Montgomery, Brandon Stewart, Shiro Kuriwaki, Chris Wlezien and participants at APSA 2023, PolMeth 2023, and the Political Methodology Speaker Series at UT Austin for their helpful comments on earlier drafts. This research was done using services provided by the OSG Consortium  (\url{https://doi.org/10.21231/906P-4D78}), which is supported by the National Science Foundation awards \#2030508 and \#2323298. Open-source software is available to implement these methods at \chref{https://github.com/mgoplerud/vglmer/tree/poisson}{https://github.com/mgoplerud/vglmer/tree/poisson}. A demonstration of how to use the accompanying software is provided in Appendix~\ref{app:software_demo}. All remaining errors are our own.}}

\newcommand{\argmax}{\operatornamewithlimits{argmax}}
\VerbatimFootnotes
\begin{document}

\maketitle 
\thispagestyle{empty}

\begin{abstract}
    Measuring public opinion at subnational geographies is critical to many theories in political science. Multilevel regression and post-stratification (MRP) is a popular tool for doing so, although existing work is limited to measuring opinion on a single survey question. We provide a framework for estimating the joint distribution of opinion on multiple questions (``Multivariate MRP''). To do so, we derive a novel method for variational inference in multinomial logistic regression with many random effects. This requires performing variational inference with high-dimensional fixed effects, but we show that this can be done at a low computational cost. We validate this procedure by estimating public opinion by party in the United States and show that existing methods can be improved considerably by adding contextual covariates on the prior levels of party identification. Substantively, we show how the output of multivariate MRP can be used to study representation across multiple policy issues simultaneously. 
    
    \vspace{0.5em}
    
    \textbf{Keywords:} Multilevel Regression and Post-stratification; Multinomial logistic regression; Variational inference; Public opinion; Policy responsiveness 
\end{abstract}

\clearpage
\pagenumbering{arabic}

\doublespacing

\section{Introduction}

Public opinion is a critical component of many theories in political science. In American politics alone, it is essential to understanding electoral behavior, substantive representation, and position-taking by candidates and parties. However, measuring attitudes among the public can be challenging, especially at the subnational level. Surveys often have too few respondents in each subnational unit to generate accurate estimates of opinion in the general population. To address this, a large literature in political science has applied various methods---most popularly multilevel regression and post-stratification (MRP; \citealt{park2004mrp})---to estimate the distribution of public opinion at subnational geographies, including states and electoral districts (\citealt{lax2009estimation,warshaw2012district}). These estimates have been used to answer substantive questions on topics as diverse as the role of public opinion in shaping state policy (\citealt{lax2012democratic}), elite misperceptions of constituent preferences (\citealt{broockman2018bias}), and class inequalities in representation (\citealt{lax2019party}). On the methodological side, research has improved MRP by using more complex hierarchical models (\citealt{ghitza2013mrp}), exploring machine learning techniques for prediction (\citealt{ornstein2019stacked,broniecki2021auto}),  speeding up estimation of the underlying models (\citealt{goplerud2023reeval}) and allowing for auxiliary, non-census, information to be included (\citealt{leemann2017extending,marble2025improving}). 

A key limitation of MRP as currently practiced is that it almost exclusively focused on estimating the distribution of one variable, either opinion from a single survey question, or a latent measure of ideology (\citealt{caughey_dynamic_2015}).\footnote{The major exception, estimating public opinion within party (e.g., \citealt{kastellec2015polarizing}), is subject of extensive discussion in Section~\ref{sec:twostep_validation}. \citet{marble2025improving} model multiple issues simultaneously but do not model the joint distribution of opinion; see Appendix~\ref{app:separate_reg} for more discussion.} Yet, a wide array of questions in political science depend on the joint distribution of public opinion across multiple questions, such as understanding how politicians take positions on several issues at a time, exploring opinion changes within parties at the subnational level, and estimating split-ticket voting from survey responses. 

To answer any of those questions at the subnational level, one must model the \emph{joint} distribution of multiple survey questions. Even with the rise of large surveys such as the Cooperative Election Study, a model-based approach is still essential for state-level estimates of public opinion on multiple issues given the lack of data when as few as two or three questions are considered. This paper provides a framework for doing so by extending MRP to obtain post-stratified estimates on the joint distribution of multiple questions in a single model  (``multivariate MRP''; \texttt{mvMRP}). We do this by using a single multinomial logistic regression whose response categories are all possible combinations of answers to the multiple questions.\footnote{For simplicity, we mostly refer to our models as multinomial logistic regression, as the standard covariates used in MRP are constant within individuals; however, our framework can accommodate covariates that differ across choices and thus includes conditional logistic regressions with hierarchical terms.} After estimation, the predictions can be post-stratified in the usual fashion and then other quantities (e.g., marginal or conditional distributions) can be easily obtained.

However, estimating multinomial logistic regression with many random effects (and possibly many response categories) is a challenging problem. It has been subject to considerable research given its importance to models in many fields, including consumer choice, transportation planning, and others (\citealt{train2009discrete}). As with other generalized linear mixed models, the key difficulty is the presence of intractable integrals over the random effects in evaluating the marginal likelihood. Early research into the topic focused on either fully Bayesian estimation or approximating the integrals using Monte Carlo methods (\citealt{train2009discrete}). More recent work has explored variational inference due to its speed on large scale problems (e.g., \citealt{braun2010discrete,tan2017stochastic,bansal2020bayesian}). Although this research tends to focus on the specific problem of ``mixed'' multinomial logistic regression (\citealt{mcfadden2000mixed}) that corresponds to a single hierarchical term, these methods are naturally extendable to the case of multiple random effects.

Unfortunately, all of these methods still inherit the problem of the intractable integrals; \cite{bansal2020bayesian} provide a comprehensive review of variational methods for mixed multinomial logistic regressions. Other approaches include stochastic variational inference to sidestep this problem entirely (\citealt{tan2017stochastic}). A different approach to high-dimensional choice---that need not be used with random effects---comes from \citet{taddy2015distributed}
who fits separate models for each response category and then combines the results to produce an aggregate prediction. While this is a potentially promising strategy, it does not allow pooling of information across different response categories which may provide better performance with limited data.

The main methodological contribution of this paper is to propose a different type of solution to avoid the issue of intractable integrals. Similar to existing work on data augmentation for variational inference in other settings (e.g., \citealt{blei2017vi}), we note that the multinomial logistic regression can be reformulated as Poisson regression with observation-specific fixed effects (e.g., \citealt{guimaraes2003tractable}). While estimating this model is usually undesirable with traditional approaches due to the addition of an extra set of high-dimensional parameters, we show that applying standard assumptions in variational inference (e.g., independence between parameter blocks) allow these parameters to be estimated at a linear cost in terms of time and memory. While this assumption is strong, a large body of work on many other problems using variational inference have shown good performance of this type of approach for recovering the posterior means of the regression parameters.

With this method in hand, we then apply it to measuring public opinion at the subnational level. Section~\ref{sec:design_MRP} discusses how to design a model for multivariate MRP by drawing connections to existing research; Section~\ref{sec:twostep_validation} validates our method on the question of estimating public opinion by party (e.g., the share of Democrats and Republicans who oppose restricting abortion access by state) as explored by a number of existing papers (e.g., \citealt{kastellec2015polarizing}). We find that multivariate MRP performs similarly to the existing two-stage approaches but naturally lends itself to more complex models that can improve performance on certain quantities of interest.

Finally, Section~\ref{sec:application} applies multivariate MRP to the study of substantive representation in the U.S. Senate. With the joint distribution of opinion on multiple issues, we extend the study of elite responsiveness to public opinion to a case with multiple issues without needing to scale opinion or roll-call votes to a single dimension of ideology. We show that senators are responsive to constituents' preferences when multiple issues are considered simultaneously, and that those representing states with fewer copartisans in the electorate are especially responsive to independent voters.

\section{Estimating Multivariate MRP}

We now consider the task of performing MRP on $J$ questions indexed $j \in \{1,\cdots, J\}$. Each question has a set of $L_j$ options whose values we note with $\mathcal{S}_j$. For example, a standard three-point party identification question in the United States would have $L_j = 3$ and $\mathcal{S}_j = \{\texttt{D}, \texttt{R}, \texttt{I}\}$, and a question about support for a policy could have $L_j=2$ and $\mathcal{S}_j = \{\texttt{yes}, \texttt{no}\}$. The goal of multivariate MRP is to obtain an estimate of the joint distribution of these questions in each subnational geography (e.g., state). In this section, we focus solely on the first step of MRP---estimating a hierarchical model on a survey---as the post-stratification procedure is unchanged from traditional MRP (see \citealt{park2004mrp}).

Assume that we have a survey with $N$ observations, indexed by $i \in \{1, \cdots, N\}$, for which we observe a response $y_i$ to all questions: $y_i = (y_{i,1}, \cdots, y_{i,J})$. Our key modelling assumption is to treat $y_i$ as a categorical variable which can take on any value in the Cartesian product of $\mathcal{S}_j$. That is, $y_i \in \bar{\mathcal{S}}$ where $\bar{\mathcal{S}} = \prod_{j=1}^J \mathcal{S}_j$ where $\bar{\mathcal{S}}$ has $\bar{L} = \prod_{j=1}^J L_j$ levels. Continuing the example from above, we combine the two variables so that $\bar{\mathcal{S}} = \{(\texttt{D}-\texttt{yes}),(\texttt{D}-\texttt{no}),(\texttt{R}-\texttt{yes}),(\texttt{R}-\texttt{no}),(\texttt{I}-\texttt{yes}),(\texttt{I}-\texttt{no})\}$. We model the probability distribution of $y_i$ as shown below with a linear predictor for each person $i$ and response category $\ell \in \bar{\mathcal{S}}$ denoted by $\psi_{i,\ell}$. This follows standard presentations for multinomial/conditional logistic regression (\citealt{agresti_categorical_2002}).

\begin{equation}
    \mathrm{Pr}(y_i = \ell) = \pi_{i,\ell} = \frac{\exp(\psi_{i,\ell})}{\sum_{\ell'=1}^{\bar{L}} \exp(\psi_{i,\ell'})}
\end{equation}

Because of the high-dimensional nature of the problem, a key insight from traditional MRP is that $\psi_{i,\ell}$ should be composed of multiple random effects for different demographic and geographic variables. The simplest formulation in the case of a single demographic variable is shown below where $\alpha_{g[i], \ell}$ denotes selecting the random effect for group $g$ of which $i$ is a member (e.g., African-American) and response category $\ell$: 

\begin{equation}
    \label{eq:simple_multinom}
    \psi_{i,\ell} = \beta_{0,\ell} + \alpha_{g[i],\ell}; \quad \alpha_{g,\ell} \sim N(0, \sigma^2_\alpha)
\end{equation}

The single random effect $\alpha_{g,\ell}$ for group $g$ and response category $\ell$ is drawn from a standard normal distribution. If the number of levels is large, it may also be useful to model the intercept $\beta_{0,\ell}$ as a random effect to avoid issues of perfect separation.\footnote{\label{ft:id}Regarding identification, it is common to set the linear predictor for some category equal to zero, e.g., $\psi_{i,1} = 0$. An alternative strategy imposes sum-to-zero constraints on the parameters. We prefer the latter for the fixed effects discussed in Section~\ref{sec:highd_fe}. For random effects, the mean-zero prior provides identification.} 

With this in hand, one could proceed to perform inference using either fully Bayesian methods (e.g., Markov Chain Monte Carlo) or alternative approaches such as REML to calibrate $\sigma^2_\alpha$ and obtain the penalized maximum likelihood estimates. However, for models used for MRP, these approaches can be prohibitively slow.

\subsection{Inference for Hierarchical Multinomial Logistic Regression}

To address this problem, we develop a novel set of inferential techniques for multinomial logistic regression with many random effects. This problem is challenging because of the intractable integrals that the random effects create when evaluating the marginal likelihood. Existing work has tackled in the multinomial logistic setting by either (i) approximating these integrals using [Monte Carlo] simulation, (ii) relying on fully Bayesian methods, or (iii) using variational inference (\citealt{bansal2020bayesian}). 

We proceed in a different direction using data augmentation to sidestep the problem of the intractable integrals entirely. Specifically, we rely on the well-known equivalence of a multinomial logistic regression as a Poisson regression with observation-specific fixed effects (e.g., \citealt{baker1994multinomial,guimaraes2003tractable}). This approach creates a new ``augmented'' dataset that has $N \times \bar{L}$ observations where each new observation is indexed by $(i,\ell)$. The outcome $\tilde{y}_{i,\ell} = I(y_i = \ell)$ is an indicator variable for whether individual $i$ chose response category $\ell$. The linear predictor $\tilde{\psi}_{i,\ell}$ corresponds to the previous linear predictor for observation $i$ and category $\ell$ plus an individual-specific fixed effect $\gamma_i$. The generative model of this Poisson representation is shown below, following the example in Equation~\ref{eq:simple_multinom}:

\begin{align}
    \label{eq:simple_poisson}
    \tilde{y}_{i,\ell} \sim \mathrm{Pois}(\lambda_{i,\ell}); \quad \lambda_{i,\ell} = \exp\left(\gamma_i + \beta_{0,\ell} + \alpha_{g[i],\ell}\right); \quad \alpha_{g,\ell} \sim N(0, \sigma^2_\alpha)
\end{align}

In the non-hierarchical case, it is known that the maximum likelihood estimates of this Poisson model correspond exactly to the estimates from the original multinomial formulation (\citealt{baker1994multinomial}). This is because if one profiled out the individual fixed effects $\gamma_i$, the profiled likelihood is identical to the original multinomial likelihood. Thus, from a fully Bayesian perspective, sampling the posterior implied the multinomial likelihood in Equation~\ref{eq:simple_multinom} and is equivalent to sampling the posterior implied by the ``augmented'' Poisson representation in Equation~\ref{eq:simple_poisson}.

A reason to prefer this reformulation is that one can use off-the-shelf software for estimating Poisson regression to fit complex multinomial/conditional logistic regression (\citealt{guimaraes2003tractable}). In the Bayesian case, however, this reformulation requires sampling a large number of extra parameters at little obvious benefit in terms of computational tractability. For the types of models used in \texttt{mvMRP}, even the original Bayesian formulation may be challenging to estimate due to there being many response categories and random effects.

A natural alternative strategy for estimation is variational inference (e.g., \citealt{grimmer2011vi,blei2017vi}). Rather than sampling the true posterior $p(\bm{\theta} | \bm{y})$, variational inference seeks to find the best approximating distribution to the true posterior---given some restrictions on the permissible approximating distribution $q(\bm{\theta})$. It is common to impose independence assumptions between blocks of parameters: In the running example, one might assume that the fixed effects, random effects, and random effect variances were all independent: $q(\bm{\theta}) = q(\bm{\beta})q(\bm{\alpha})q(\sigma^2_\alpha)$. It is sometimes necessary to impose distributional assumptions, e.g., $q(\bm{\beta})$ and $q(\bm{\alpha})$ are multivariate Gaussian. It can be shown that estimation using variational inference reduces to maximizing the following objective function---known as the evidence lower bound (ELBO)---over all permissible variational distributions $q(\bm{\theta})$. Equation~\ref{eq:VI} shows the objective, where $\ln p(\bm{y}, \bm{\theta})$ denotes the sum of the log-likelihood of the data $\bm{y}$ given the parameters $\bm{\theta}$ and the log-prior on all parameters $\bm{\theta}$:

\begin{equation}
    \label{eq:VI}
    q^*(\bm{\theta}) = \argmax_{q(\bm{\theta})}~ \mathrm{ELBO}_{q(\bm{\theta})}; \quad \mathrm{ELBO}_{q(\bm{\theta})} = \mathrm{E}_{q(\bm{\theta})}\left[\ln p(\bm{y}, \bm{\theta}) - \ln q(\bm{\theta})\right]
\end{equation}

Performing variational inference with a multinomial likelihood is challenging, however, as even assuming $q(\bm{\beta})$ and $q(\bm{\alpha})$ are Gaussian is insufficient to create a tractable ELBO given the presence of integrals that must be evaluated numerically. The key challenge is the ``log-sum-exp'' term in the denominator of the multinomial likelihood: $\ln\left(\sum_{\ell'=1}^{\bar{L}} \exp(\psi_{i,\ell'})\right)$.

To sidestep this, authors have proposed maximizing some lower bound or approximation to the ELBO (e.g., \citealt{braun2010discrete}). These result in a tractable model but are likely worse approximations as they are not targeting the true objective function. Other alternatives exist, such as using simulation to approximate the integrals (\citealt{bansal2020bayesian}) or stochastic variational inference (\citealt{tan2017stochastic}). 

In this paper, we use the alternative representation of the discrete choice problem as a Poisson likelihood to obtain a tractable problem. Using the posterior implied by Equation~\ref{eq:simple_poisson}, a mean-field assumption between the fixed effects $q(\bm{\gamma})$ and the main parameters of interest $q(\bm{\theta})$, and standard assumptions (i.e., $q(\bm{\theta}) = q(\bm{\beta})q(\bm{\alpha})q(\sigma^2_\alpha)$ and Gaussian distributions on $\bm{\beta}$, $\bm{\alpha}$, and $\bm{\gamma}$) the ELBO can be evaluated in closed-form. This builds on the tradition of using data augmentation and variational inference where the distributions of the augmentation variables are assumed to be independent of the main parameters of interest (\citealt{blei2017vi}). Such an approach has shown good performance for other generalized linear mixed models (e.g., logistic regression; \citealt{goplerud2022mavb}).

Equation~\ref{eq:elbo_poisson} shows this augmented variational problem. In theory, both $q^*(\bm{\theta})$ and $q^*(\bm{\gamma})$ can be estimated using coordinate ascent variational inference (CAVI) without requiring further assumptions or lower-bounding of the ELBO.

\begin{equation}
    \label{eq:elbo_poisson}
    q^*(\bm{\theta}), q^*(\bm{\gamma}) = \argmax_{q(\bm{\theta}), q(\bm{\gamma})}~ \mathrm{E}_{q(\bm{\theta}), q(\bm{\gamma})}\left[\ln p(\bm{y}, \bm{\theta}, \bm{\gamma})\right] - \mathrm{E}_{q(\bm{\theta})}\left[\ln q(\bm{\theta})\right] - \mathrm{E}_{q(\bm{\gamma})}\left[\ln q(\bm{\theta})\right]
\end{equation}

In practice, however, estimating the optimal $q^*(\bm{\theta})$ and $q^*(\bm{\gamma})$, requires care given the large dimensionality of $\bm{\gamma}$ with one parameter for each of the $N$ respondents. In the next two sub-sections, we outline our novel approach to estimating high-dimensional fixed effects for variational inference as well as how to deal with the Poisson likelihood itself.

\subsubsection{Variational Inference with High-Dimensional Fixed Effects}\label{sec:highd_fe}

The first issue is around the high-dimensional fixed effect $\bm{\gamma}$ that consists of $N$ parameters. Even with the independence assumption from $q(\bm{\theta})$, inference is challenging. The key issue arises from the fact that $\bm{\gamma}$ must be constrained or modified to identify the model. Two strategies are popular: First, one might set a reference category where some $\gamma_i = 0$ and define all fixed effects as the difference from $\gamma_i$. We prefer a different, equivalent, approach of constraining all $\bm{\gamma}$ to sum-to-zero, e.g., $\bm{1}^T\bm{\gamma} = 0$. This is more general insofar as it allows multiple fixed effects to be included, avoids keeping track of an (arbitrary) reference category, and would allow for more complex constraints to be imposed.

Appendix~\ref{app:VI_FE} derives a result, which we believe to be novel, about the optimal variational approximation $q^*(\bm{\gamma})$ for a multivariate Gaussian distribution given linear equality constraints $\bm{L}\bm{\gamma} = \bm{0}$. However, in our application, the optimal approximation $q^*(\bm{\gamma})$ has a dense $N \times N$ covariance matrix. Thus, naively computing and using $q^*(\bm{\gamma})$ would incur large costs in terms of memory and computing time. An additional result in Appendix~\ref{app:VI_FE} shows that, for the design matrix and constraint used when estimating high-dimensional fixed effects, the relevant terms of $q^*(\bm{\gamma})$ needed to evaluate the ELBO and obtain $q^*(\bm{\theta})$ can be computed extremely inexpensively---in linear time and memory. As a result, our approach of turning a multinomial regression into an augmented Poisson regression is feasible even if $N$ is enormous. This method can be immediately applied beyond our specific problem to estimating fixed effects using variational inference in other settings.

\subsubsection{Variational Inference with Poisson Likelihoods}

We also must perform variational inference on a Poisson likelihood. There are a number of popular approaches to do this; most naturally, note that if we assume that the fixed and random effects have a normal distribution, the objective function (ELBO) can be evaluated exactly given that $\mathrm{E}\left[\exp(x)\right] = \exp(\mu + \frac{1}{2} \sigma^2)$ if $x \sim N(\mu, \sigma^2)$. \cite{wand2014fully} proposes a non-conjugate variational message passing algorithm to update those parameters and provides a straightforward algorithm for estimating a hierarchical model with a Poisson likelihood. Appendix~\ref{app:VI_poisson} provides a full derivation of the algorithm.

\subsection{Inference Using Separate Regressions}\label{sec:separate_reg}

An important potential alternative methodology for predicting categorical outcomes consists of fitting separate regressions and combining the results into a single predictive model. \cite{taddy2015distributed} provides a theoretical justification for doing so based on using a plug-in estimator of the fixed effect $\gamma_i$. The intuition is that if $\gamma_i$ were known, then the multinomial logistic likelihood ``separates'' and can be estimated using $\bar{L}$ entirely separate Poisson regressions. In our setting, \citeauthor{taddy2015distributed}'s (\citeyear{taddy2015distributed}) plug-in estimator $\hat{\gamma}_i=0$ for all $i$; thus, his proposal corresponds to simply ignoring the fixed effects entirely and fitting $\bar{L}$ separate (hierarchical) Poisson regressions. Prediction is done by normalizing the expected outcomes across categories. Appendix~\ref{app:separate_reg} provides a detailed discussion of how this separate regressions can be used as well as drawing connections to the popular ``One-vs-All'' approach from machine learning.

A downside of these approaches, however, is that they do not allow information to be pooled across response categories as the models are separately estimated. Further, covariates that are constant within a response category for all observations cannot be included. Given the limited data available for MRP, we thus expect this approach to do less well. An intermediate approach is possible by borrowing insights from random effects: Instead of $\bar{L}$ separate regressions, one could estimate them simultaneously and partially pool the intercepts across the models (see Appendix~\ref{app:separate_reg} for more details and \citealt{marble2025improving} for a related idea). Interestingly, in practice, this would look identical the proposed \texttt{mvMRP} model above but (i) omitting the fixed effects and (ii) re-normalizing the predictions appropriately. These partially pooled models can thus be easily fit using the same inferential framework described in this paper and in the accompanying software.

\section{Designing Models for Multivariate MRP}\label{sec:design_MRP}

This section outlines some principles for designing a model for multivariate MRP that fit into the above generic framework for inference on hierarchical multinomial logistic regression. 

We make this concrete by focusing on a two-question example: Consider the joint estimation of \texttt{partyID} (party identification)---for which valid answers are \texttt{D}, \texttt{R}, and \texttt{I}---and \texttt{abortion} (banning abortion after 20 weeks)---for which the valid answers are \texttt{support} or \texttt{oppose}. Continuing with the earlier example, we assume that there is one demographic variable (\texttt{race}), and the geographic unit of interest is \texttt{state}, for which we have one contextual covariate of \texttt{demvote} (i.e., the vote share for the Democratic candidate for president at the last election; \citealt{buttice2013mrp}). 

When designing a multivariate MRP model, it is important to recall that $\ell$ corresponds to some particular combination of choices across all questions, e.g., $\ell = (\texttt{D}-\texttt{oppose})$. We use $\ell_{\texttt{partyID}}$ to refer to the choice of party identification associated with $\ell$ (e.g., \texttt{D}) and $\ell_{\texttt{abortion}}$ to refer to the choice associated with the abortion question (e.g., \texttt{oppose}).

It can be helpful to think about $\psi_{i, \ell}$ as representing the systematic ``utility'' of that choice; one can conceptualize a multinomial logistic regression as an individual $i$ choosing the choice $\psi_{i, \ell}$ with the highest utility---subject to i.i.d. random noise added to each $\psi_{i,\ell}$ (\citealt{mcfadden_conditional_1974,train2009discrete}). We proceed by discussing how to include traditional variables for MRP as well as new types of variables that may be especially useful for multivariate MRP.

\textbf{Traditional MRP Variables:} All of the ``standard'' variables used in MRP (demographics, geography, and contextual covariates) are fixed within an individual. Thus, they must \emph{interact} with the questions or else they cannot be included in the underlying model. This is standard for all multinomial logistic regressions and appears as having a coefficient on, say, race that varies by response category (\citealt{agresti_categorical_2002}).

Given the high-dimensionality of the problem resulting from the ``intersection'' of multiple questions, it is useful to ensure that the model has---at least---a simple additive structure. We illustrate this below where the linear predictor $\psi_{i, \ell}$ as consisting of four additive terms: (i-ii) an ``intercept'' for each question of \texttt{partyID} and \texttt{abortion} and (iii-iv) an effect of race for each question.

\begin{equation}
    \psi_{i,\ell} = \beta^{\texttt{partyID}}_{0,g[\ell_{\texttt{partyID}}]} + \beta^{\texttt{abortion}}_{0,g[\ell_{\texttt{abortion}}]} + \alpha^{\texttt{partyID}}_{g[i], g'[\ell_{\texttt{partyID}}]} + \alpha^{\texttt{abortion}}_{g[i], g'[\ell_{\texttt{abortion}}]}
\end{equation}

Another implication of this simple additive formulation is that multivariate MRP is inherently ``deep'' (\citealt{ghitza2013mrp}), insofar as all random effects for demographics and geography are implicitly interacted with the choice ($\ell$) or some constituent question ($\ell_j$). 

As the mathematical notation to explain the model can get rather cumbersome given the number of interactions, it is helpful to illustrate the model using the popular \texttt{lme4} syntax for estimating random effects (\citealt{bates2015lmer}). Our accompanying software is written to accept this syntax, see Appendix~\ref{app:software_demo} for an illustration. In this notation, the above model can be expressed as follows, where \verb@y~x+(1|g)@ would indicate a ``fixed effect'' on \texttt{x} and a random intercept for group \texttt{g}. The notation \verb@(1 | a : b)@ indicates a random intercept on the interaction of variables \texttt{a} and \texttt{b}.

\begin{verbatim}
    response ~  (1 | partyID) + (1 | abortion) + 
                (1 | partyID : race) + (1 | abortion : race)
\end{verbatim}

As existing work shows (\citealt{ghitza2013mrp,goplerud2023reeval}), it is likely useful with sufficient data to go ``deeper'' and estimate models with complex interactions. In this case, one could make the model richer by (a) adding interactions between demographics and questions, [e.g., \verb@(1 | race : sex : partyID )@] or (b) between questions [e.g., \verb@(1 | partyID : abortion)@ or \verb@(1 | race: partyID : abortion)@]. Despite the resulting complexity of these deep models, they can be estimated rapidly using the variational methods proposed above. 

Another critical part of the effectiveness of MRP is the role of ``contextual'' covariates, i.e., covariates known at the level of geography to which one is post-stratifying, such as presidential vote share (\citealt{lax2009estimation,buttice2013mrp}). These covariates are known to provide large improvements in performance in MRP, and we expect them to also have a critical role here.

However, their inclusion is more complex in that an especially good contextual covariate should well predict \emph{all} response categories. This may often require using different contextual covariates for different questions; a statewide variable that does a good job at predicting abortion attitudes may be different from one that predicts party identification. Section~\ref{sec:twostep_validation} suggests some choices for these covariates. 

Once those contextual covariates have been selected, entering them into the model is straightforward. As with the demographics, $x_i$ has the same value for all choices $\ell$ and must be interacted with a response category or question to be included in the model. One approach is a random slope, shown below:

\begin{equation}
    \psi_{i,\ell} = \cdots + \gamma^{\texttt{partyID}}_{1,g[\ell_{\texttt{partyID}}]} x_{i} + \cdots; \quad \left[\begin{array}{l} \beta^{\texttt{partyID}}_{0,g} \vspace{0.5em} \\  \gamma^{\texttt{partyID}}_{1,g}
    \end{array}\right] \sim N\left(\bm{0}, \bm{\Sigma}_{\texttt{partyID}}\right)
\end{equation}

In \texttt{lme4} syntax, these can be specified as \verb@(1 + x | partyID)@.

\textbf{Choice-Specific Covariates:} The above discussion focuses on the ``pure'' multinomial logistic regression where all covariates are fixed within a respondent, i.e. constant across all $\psi_{i,\ell}$. A different type of covariate---alternative or choice-specific covariate---takes on different values across $\ell$ (\citealt{agresti_categorical_2002}). Models with these covariates are often known as ``conditional logistic regression'' (\citealt{agresti_categorical_2002}). We also note that our representation of the data in the ``expanded'' Poisson formulation is identical to how data is prepared for conditional logistic regression.

As we illustrate in Section~\ref{sec:twostep_validation}, such covariates can be quite useful for \texttt{mvMRP}. We focus on the role of lagged copartisanship (i.e., the share of copartisans in a state measured from a previous survey) given our focus on party identification and policy. For example, if $\ell=(\texttt{D}-\texttt{oppose})$, one would include the lagged share of Democrats as an alternative-specific covariate. If one were analyzing, say, vote choice for candidates, a measure of incumbency or candidate quality would likely be a highly important alternative-specific covariate.

In terms of including it in the model, this covariate is now defined by $x_{i,\ell}$ (as it varies across choice $\ell$) and thus can be included directly in the linear predictor without interactions, e.g., $\beta x_{i, \ell}$, but could be interacted with question(s) if desired.

\subsection{Quantities of Interest}
\label{sec:mrp_qoi}

Multivariate MRP opens up a new array of quantities of interest to be calculated after the fact. We suggest that one first calculates the joint distribution of public opinion across questions at the relevant geographic level---e.g. the proportion of voters in a state who are Democrats and oppose restricting abortion access. From there, one can back out various other quantities. Following \cite{kastellec2015polarizing}, one might be interested in conditional probabilities: What proportion of Democrats support restricting abortion access? Appendix~\ref{app:responsivness_qoi} outlines some additional quantities (e.g., heterogeneity within party) that one might compute. Additionally, we note that one can also back out the ``usual'' univariate MRP estimates from this model: the distribution of party identification in each state or the support for a policy across \emph{all} voters. These are often times more straightforward to validate and should be examined to ensure that the multivariate model has done a reasonable job capturing more simple quantities of interest.

\section{Validating Multivariate MRP}
\label{sec:twostep_validation}

We validate our method against the closest existing alternative: ``Two-Stage MRP'' (\citealt{kastellec2015polarizing}), which estimates public opinion by party inside of each state. For example, what percentage of Democrats in Pennsylvania support abortion restrictions? As this procedure is less commonly used than traditional MRP, we explain the logic here and draw connections to multivariate MRP.

\subsection{Two-Stage MRP}

The ideal approach to estimating opinion at the state-party level would simply be to add adding respondents' party identification as a predictor to the ``standard'' MRP model and likewise including it in the post-stratification stage. Using the \texttt{lme4} syntax, such a model could take the form:
\begin{verbatim}
    response ~ (1 | party) + (1 | age) + (1 | educ) + (1 | state) + ...
\end{verbatim}
However, this is generally not possible in MRP, as party identification is not available in census data and therefore cannot be used in the post-stratification step. \citet{kastellec2015polarizing} propose resolving this problem by modeling opinion in two stages, also known as MRP for non-census variables (\citealt{lopezmartin2022multilevel}).

First, they fit a hierarchical model to predict party identification---usually a three-level variable (\texttt{D}, \texttt{R}, \texttt{I})---using the standard MRP variables:
\begin{verbatim}
    party ~ (1 | age) + (1 | educ) + (1 | race) + (1 | state) + ...
\end{verbatim}
This model is post-stratified to predict party identification using census variables, thereby extending the typical variables available for post-stratification to include (modeled) party.

Then, in the second stage, they fit the ideal model of opinion described above, including party, and perform post-stratification using the probabilities of demographics and party identification obtained from the first-stage model. Formally, given a set of predictors $\bm{z}_i$, one calculates the joint probability of an outcome as follows:
\begin{align}
\Pr\left(\begin{aligned}\texttt{policy}_i &= \ell', \\ \texttt{partyID}_i &= \ell \end{aligned} \;\middle|\; \bm{z}_i\right) = \Pr(\texttt{policy}_i = \ell' | ~\texttt{partyID}_i = \ell, \bm{z}_i) \times \Pr(\texttt{partyID}_i = \ell | \bm{z}_i)
\end{align}
Appendix~\ref{app:twostage_equiv} shows that, in a simplified setting, a two-stage approach of this form is, in fact, an approximation of a multinomial logistic regression where interaction terms are omitted.

While this is a reasonable procedure, Two-Stage MRP has some important limitations that our framework addresses. First, while this works reasonably well for two questions, it is challenging to see how to extend this to more than two questions. One would need to specify a series of conditional models to build to the full joint distribution. The order of these models is likely non-trivial to choose systematically and may affect the ultimate estimates. Second, it is difficult to propagate uncertainty fully through the two-stage model. By contrast, a single multinomial model can directly propagate uncertainty to all estimated quantities.\footnote{We do not consider uncertainty in this paper as it is generally poorly captured by mean-field variational inference. If essential, one could bootstrap the variational estimates to roughly quantify uncertainty (\citealt{imai2016ideal}).} We also note that the first stage of this procedure involves modeling a categorical variable, and thus our methodology could be used here as well. Indeed, (non-fully Bayesian) inference multinomial regression with many random effects is rather challenging with existing software; Appendix~\ref{app:software_other} provides a review.

\subsection{Comparing Multivariate MRP and Two-Stage MRP}

To validate multivariate MRP, we follow a common strategy for validating MRP more generally of making a ``superpoll'' that pools together many surveys with identical question wording to create an estimate of the ground truth (e.g., \citealt{buttice2013mrp}). This procedure takes a small sample of this dataset, performs MRP, and compares the estimates against the ``truth'' from the superpoll. For each policy, our goal is to estimate the distribution of opinion and three-point party identification (Democrat, Republican and Independent) in each state. 

To validate this method systematically, we need to have a larger superpoll than in existing approaches given that we are trying to estimate the joint distribution of policy and party within each state. To do this, we looked for issues that were included on the Cooperative Election Study (CES) with stable question wording in three consecutive waves (2016-2018-2020 or 2014-2016-2018). We located six questions, briefly described below:

\begin{itemize}
    \item Abortion (20 Weeks): Support for banning abortion after the 20th week of pregnancy
    \item Abortion (Ban): Support for making abortion illegal in all circumstances
    \item Assault Weapons: Support for banning assault rifles
    \item Gay Marriage: Support for allowing gays and lesbians to marry
    \item Immigration: Support for granting legal status to certain undocumented immigrants
    \item Renewable Energy: Support for requiring states to use a minimum amount of renewable energy
\end{itemize}

This gives us around 170,000 respondents for each policy, with a median of around 750 respondents for each state/party combination. We drew samples of 2,000 respondents from the combined poll and then compared the MRP estimates against the true value from the superpoll. Appendix~\ref{app:validation_CCES} provides additional information about the survey and MRP procedure, including the exact question wording and details about the construction of the post-stratification weights.
 
In each MRP model, we rely on standard contextual covariates for social issues (democratic vote share for president at the prior election and the percent of evangelicals in each state; \citealt{buttice2013mrp}). We used the following demographic predictors mirroring typical validations of MRP: age, education, sex, and race. To estimate a model that mirrors the additive specification above, we include a random intercept for the interaction between each demographic variable and question/response category, an indicator variable for the response category, and random slopes for each contextual covariate by question/response category. We also include random effects for state and Census division, again interacted with question/response category. Appendix~\ref{app:software_demo} contains the full formula and code for estimating this in the accompanying software.

We first note that this is a rather complex model even for a simple additive case of multivariate MRP: It contains 21 random effects and thus estimating this model without using variational methods is likely quite expensive. With variational inference, however, estimation takes around three minutes on a remote server with 8GB of RAM.

\begin{figure}[!htbp]
\caption{Marginal Distribution of Partisanship}
\vspace{-1em}
\label{fig:partisan}
\includegraphics[width=0.9\textwidth]{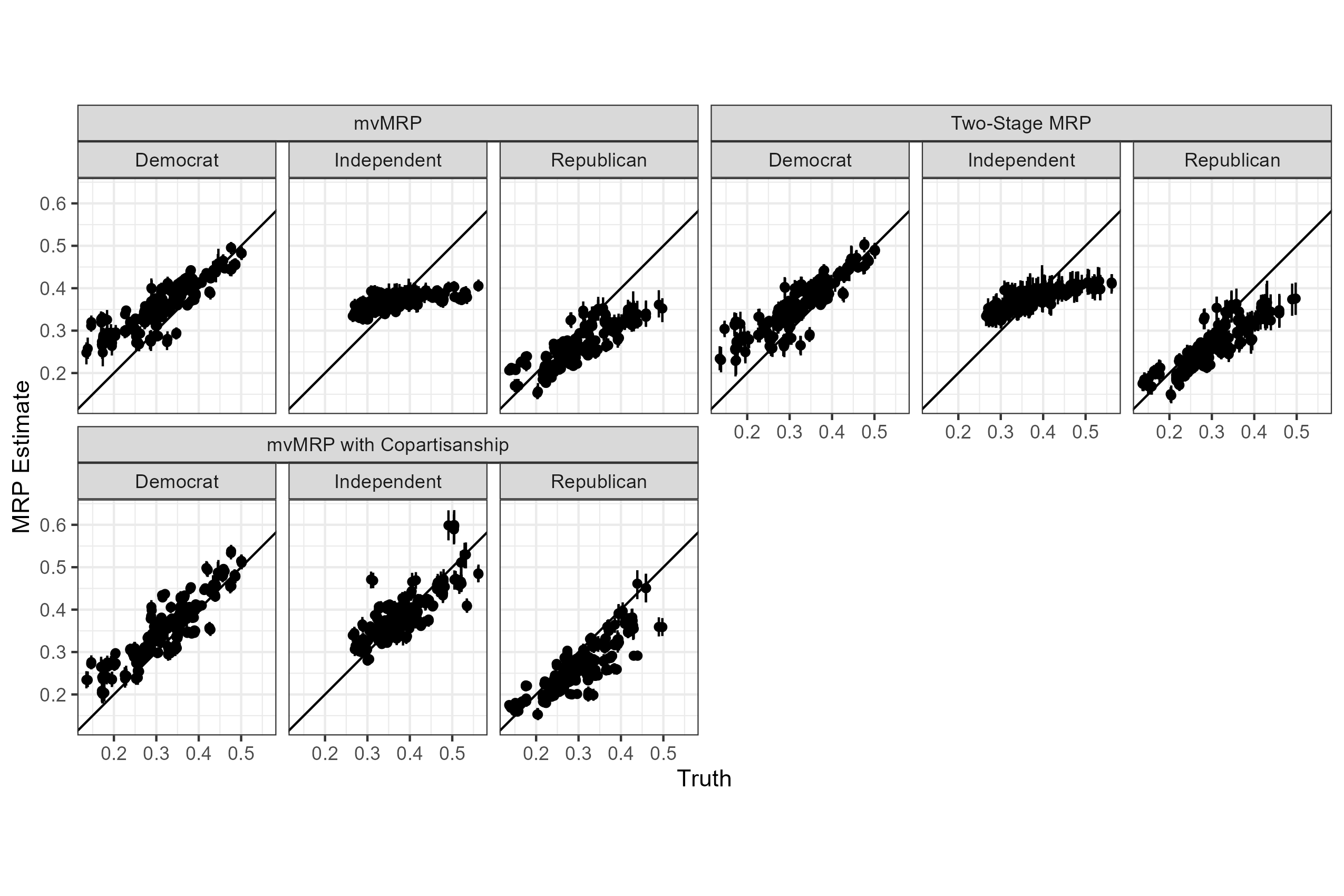}
\caption*{\footnotesize \emph{Note}: This figure shows the estimates of the marginal distribution of partisanship in each state for each policy, averaged across 500 simulations. The 25th and 75th quantiles are shown. ``mvMRP with Copartisanship'' refers to the model that includes lagged copartisanship.}
\end{figure}

We also considered Two-Stage MRP using \texttt{mgcv} as it is one of the few packages in \texttt{R} to estimate multinomial logistic regression with random effects (see Appendix~\ref{app:software_mgcv} for the exact specification). When inspecting these preliminary models, it was evident that both Two-Stage MRP and multivariate MRP performed quite badly in terms of estimating the distribution of party identification. Figure~\ref{fig:partisan} shows the average estimated value of partisanship for each stage---i.e. marginalizing out the policy question from the joint distribution---across 500 simulations. It is clear from the top two panels that the share of independents is poorly estimated as it appears too flat versus the truth.

To address this, we included an alternative-specific contextual covariate: lagged copartsianship. This is a useful predictor insofar as the ``baseline'' level of party identification in a state is likely closely correlated with its past value. This also addresses a weakness of traditional Two-Stage MRP as the commonly used contextual covariates (\texttt{demvote}) have very little predictive power over the probability that a respondent is \emph{independent}---while it does a good job of predicting the share of Republicans and Democrats. We calculate this by using the lagged disaggregated estimate from the prior CES.\footnote{Where this is unavailable, one might perform \texttt{mvMRP} with a categorical outcome to predict partisanship.}

We thus created a second version of multivariate MRP---``Copartisan \texttt{mvMRP}''---that added this as an unregularized term in the linear predictor, as well as having random slopes on its effect by question and response category. As Figure~\ref{fig:partisan} shows, this immediately improved the issue considerably---although some outliers still remain. This model improved the error on estimating the share of independents in a state by around 25\% on average, as well as improving the error on the share of Democrats (13\% improvement) and Republicans (9\%).

We benchmark our approach against some additional methods. First, we implement a ``naive'' approach that fits two separate (traditional) MRP models---one to predict partisanship and one to predict the policy question. We then combine these together assuming the responses are independent given respondent covariates. This is obviously a poor model for joint and conditional probabilities, but it is a more parsimonious model for the marginal distributions as these are estimated separately. It thus provides some evidence as to whether Two-Stage MRP or multivariate MRP can improve upon the estimation of the marginal distributions of opinion. We also implement \cite{taddy2015distributed}'s approach of fully separate Poisson regressions with gamma lasso regression (``Taddy'').

We explore performance on four relevant quantities: (i) the joint distribution of opinion at the state-level (e.g., the share of individuals who are [a] Democratic and [b] support the policy), (ii) the conditional distribution by party (i.e., the support for a policy by party in each state), and (iii-iv) the marginal distribution of party identification and policy response. In all of these results, we compute the mean absolute error (MAE) in the predicted quantity against the truth, averaged across all fifty states. To address variability in the 2,000-person sample chosen, we report $\texttt{MAE}_{p,k}$ for policy $p$ and method $k$ having taken the median over 500 simulations.

Table~\ref{tab:validate_joint} begins by showing the results for quantities that depend on both questions--the joint distribution and conditional distribution of support by party. To present the results in an interpretable fashion, we compute the percentage change in MAE versus Two-Stage MRP (similar to \citealt{broniecki2021auto}). Each cell in the table reports $\left(\texttt{MAE}_{p,k} - \texttt{MAE}_{p,\texttt{Two-Stage}}\right)/\texttt{MAE}_{p,\texttt{Two-Stage}} \cdot 100$ where negative numbers indicate that method $k$ out-performs Two-Stage MRP on policy $p$. Appendix~\ref{app:validation_raw_mae} reports the raw value of the mean absolute error. To summarize across policies, the final row, in bold, reports the average percentage improvement across the six policies.

\begin{table}
\caption{Validation of Joint and Conditional Distributions}
\label{tab:validate_joint}
\begin{centering}
\begin{tabular}{r@{\,}r|r|r|r|r}
\hline\hline
 & \multicolumn{1}{c|}{Policy} & Naive & Taddy & \texttt{mvMRP} & Copart. \\
\ldelim\{{7}{*}[Joint~] & Abortion (Ban) & 6.73 & 26.93 & 0.27 & -8.04 \\
 & Abortion (20 Weeks) & 67.79 & 28.50 & -0.51 & -12.42 \\
 & Assault Weapons & 88.64 & 28.79 & 1.09 & -7.32 \\
 & Gay Marriage & 70.12 & 24.43 & 1.96 & -7.93 \\
 & Immigration & 56.75 & 21.41 & -1.04 & -12.24 \\
 & Renewable Energy & 67.08 & 23.19 & 0.21 & -9.60 \\
 & \textbf{Average} & \textbf{59.52} & \textbf{25.54} & \textbf{0.33} & \textbf{-9.59} \\
\ldelim\{{7}{*}[Conditional~] & Abortion (Ban) & 107.05 & 24.16 & 0.44 & 0.43 \\
 & Abortion (20 Weeks) & 228.12 & 20.19 & -4.53 & -4.47 \\
 & Assault Weapons & 303.97 & 32.46 & 0.63 & 2.04 \\
 & Gay Marriage & 230.63 & 38.48 & -1.86 & -1.96 \\
 & Immigration & 216.62 & -0.50 & -5.00 & -4.93 \\
 & Renewable Energy & 234.52 & 10.94 & -3.60 & -3.44 \\
 & \textbf{Average} & \textbf{220.15} & \textbf{20.95} & \textbf{-2.32} & \textbf{-2.05}
\\\hline\hline
\end{tabular}
\caption*{\footnotesize \emph{Note:} This table shows the percentage change in mean absolute error across states. Negative numbers indicate the method out-performs Two-Stage MRP. The first seven rows show the error on estimating joint opinion; the latter seven rows show the error on estimating opinion conditional party. The models are described in the main text where ``Copart.'' is the \texttt{mvMRP} model that includes lagged copartianship.}
\end{centering}
\end{table}

Table~\ref{tab:validate_joint} shows promising results for \texttt{mvMRP}. In the model without copartisanship, it is generally within a percentage point of Two-Stage MRP for estimating joint opinion and around 2\% better for estimating conditional opinion on average---although some policies show considerable gains.

When lagged copartisanship is included with \texttt{mvMRP} (``Copart.''), the performance increases dramatically for estimation of joint opinion---improving by nearly 10\% on average and by at least 7\% on all issues. To contextualize this result, the improvement that one finds from using machine learning or a deep model is often around 5\% to 12\% (e.g., \citealt{broniecki2021auto,goplerud2023reeval}). Put another way, across all simulations and issues, \texttt{mvMRP} with copartisanship decreases the mean absolute error versus Two-Stage MRP in around 97\% of simulations for estimating joint opinion and 60\% for conditional opinion. 

As expected, the naive method performs extremely poorly for estimating joint or conditional distributions---doing around 60\% or 200\% worse on average, respectively. More interestingly, fully separate regressions (``Separate'') perform also rather poorly doing about 20\% worse than Two-Stage MRP on average. Appendix~\ref{app:validation_separate} explores this in more detail; it finds that partially pooling the models using random effects (i.e., estimating the \texttt{mvMRP} models without the high-dimensional fixed effect; see Section~\ref{sec:separate_reg}) returns close performance to \texttt{mvMRP}. Thus, for the size of surveys (1,000-2,000 respondents) for which MRP is often performed, it appears essential to partially pool across response categories.

Turning next to the marginal quantities, Table~\ref{tab:validate_marginal} shows that, as expected, if the quantity of interest is the marginal distribution of partisanship, one can do considerably better if lagged co-partisanship is included (around a 15\% improvement); this is also true for the naive model---as its marginal model for party includes this covariate. 

\begin{table}
\caption{Validation of Marginal Distributions}
\label{tab:validate_marginal}
\begin{centering}
\begin{tabular}{r@{\,}r|r|r|r|r}
\hline\hline
 & \multicolumn{1}{c|}{Policy} & Naive & Taddy & \texttt{mvMRP} & Copart. \\
\ldelim\{{7}{*}[\shortstack{Party\\Marginal}~] & Abortion (Ban) & -12.70 & 27.41 & -0.09 & -13.66 \\
 & Abortion (20 Weeks) & -14.18 & 27.70 & -0.03 & -16.14 \\
 & Assault Weapons & -12.28 & 24.97 & -0.53 & -15.32 \\
 & Gay Marriage & -14.19 & 15.17 & 3.45 & -14.76 \\
 & Immigration & -14.85 & 25.96 & 0.74 & -15.61 \\
 & Renewable Energy & -13.23 & 26.21 & 0.86 & -14.52 \\
 & \textbf{Average} & \textbf{-13.57} & \textbf{24.57} & \textbf{0.73} & \textbf{-15.00} \\
\ldelim\{{7}{*}[\shortstack{Policy\\Marginal}~] & Abortion (Ban) & -1.11 & 44.18 & 7.91 & 7.99 \\
 & Abortion (20 Weeks) & -3.58 & 71.24 & -1.13 & 1.07 \\
 & Assault Weapons & -4.85 & 82.86 & 7.91 & 1.86 \\
 & Gay Marriage & -1.55 & 69.87 & 4.56 & 8.79 \\
 & Immigration & 0.12 & 25.45 & -6.23 & -5.38 \\
 & Renewable Energy & -4.14 & 39.45 & 1.21 & -2.45 \\
 & \textbf{Average} & \textbf{-2.52} & \textbf{55.51} & \textbf{2.37} & \textbf{1.98}
\\\hline\hline
\end{tabular}
\caption*{\footnotesize \emph{Note:} This table shows the percentage change in mean absolute error across states as discussed in the main text. Negative numbers indicate the method out-performs Two-Stage MRP. The first seven rows show the error on estimating the marginal distribution of partisanship; the latter seven rows show the error on marginal distribution of the policy. The models are described in the main text and Table~\ref{tab:validate_joint}.}
\end{centering}
\end{table}

If the goal is estimating marginal opinion on policy, however, the simulations suggest that---in fact---one should simply use the ``naive'' method as this performs about 2.5\% better than Two-Stage MRP. This is sensible; one uses a simpler model for a simple quantity and, given limited data, one sees better performance. We see that \texttt{mvMRP} (both variants) do slightly worse on average than Two-Stage MRP for estimating this quantity, with considerably more variability across issues than other quantities of interest. Appendix~\ref{app:validation_mean_vs_median} shows, however, that if one uses the average MAE across simulations (vs. the median MAE), \texttt{mvMRP} does slightly better than Two-Stage MRP; this is because of occasional simulations where \texttt{mgcv} does not regularize the random effects adequately.

Appendix~\ref{app:validation} contains some additional results. Appendix~\ref{app:validation_copart} explores including copartisanship into Two-Stage MRP as well as estimating both stages using the variational inference methodology proposed by this paper. It finds generally comparable performance, although \texttt{mvMRP} shows larger gains to estimating the joint distribution. Appendix~\ref{app:validation_shallow} provides additional simulation results on different specifications of \texttt{mvMRP}, examining both a simpler model and one with more interactions. Both perform comparably or slightly worse than the specification in the main text.

\section{Exploring Multidimensional Responsiveness}
\label{sec:application}

Our final application of multivariate MRP uses it to tackle a thorny question about democratic representation. Scholars have routinely found that politicians are responsive to the preferences of their constituents; that is, that public opinion is predictive of legislators' votes on policy (e.g., \citealt{erikson1978constituency,page1983effects,lax2012democratic,caughey_dynamic_2015}). Studies in this tradition have generally examined elite position-taking in one of two ways. Many studies consider policies on an issue-by-issue basis, which assumes that politicians consider issues independently. Alternatively, scholars may scale public opinion and legislator roll-call votes into a single dimension of ideology. However, this may mask important variation in opinion by confounding moderates and those with extreme but idiosyncratic preferences (\citealt{broockman2016approaches}).

Yet, it is well established that the electorate often holds complicated and cross-pressured preferences that may not be well captured by examining one issue at a time, or by scaling ideology (\citealt{achen1978measuring,dahl1956preface}, p. 126). Likewise, reelection-minded legislators are strategic in their behavior, considering public opinion across multiple issues at a time and also taking into account the salience of issues and preferences of relevant subgroups of their constituencies (\citealt{arnold1990logic}). A natural extension of the literature on elite responsiveness is to consider representation as a multidimensional concept---that is, looking at legislators' decisions on multiple issues at once. In order to do so, it is necessary to understand the joint distribution of public opinion across several issue dimensions at once. Multivariate MRP allows us to do this.

\subsection{Context and Data}

This section presents an initial example using \citet{lax2019party}'s data on votes in the U.S. Senate. To illustrate the utility of multivariate MRP, we selected a pair of issues for which there was some disagreement within each party on the roll-call votes. Both votes were taken during the 113th Congress and are briefly summarized below:

\begin{itemize}
    \item Assault Weapons Ban: This proposal, which failed to pass the Senate (40-60; 38-15 for Democrats and 1-44 for Republicans), would have prohibited the manufacture and sale of more than 150 assault rifles and other firearms, as well as high-capacity magazines. 
    \item Farm Bill (Agricultural Act of 2014): This proposal, enacted into law, passed the Senate by a 68-32 vote (44-9 for Democrats; 22-23 for Republicans). It reauthorized (but cut) agricultural subsidies as well as cutting food stamps. 
\end{itemize}

On both bills, there was considerable disagreement among senators within party. While the modal vote for Democrats (29) was to vote ``yes'' on both bills, and the modal vote for Republicans (23) was to vote ``no'' on both, 44 senators cast a different, more complex pattern of votes. Appendix~\ref{app:responsiveness_bills_surveys} lists the full distribution of senators' positions.

To examine this variation further, we match senators' roll-call votes on these issues with opinion estimates from \texttt{mvMRP}, discussed below. We then consider four hypotheses about how public opinion shapes senators' votes across issues. First, we explore whether greater statewide support for a particular combination of policies (e.g., Yes on Farm Bill, No on Assault Ban) corresponds to senators' likelihood of casting the corresponding votes (Hypothesis 1). This is a natural extension of typical responsiveness results (e.g., \citealt{erikson1978constituency}) to the multidimensional case.

Next, we test existing hypotheses that suggest senators may be more responsive to the policy preferences of some constituents. On the one hand, some literature (e.g., \citealt{wright1989policy}) expects more responsiveness to the preferences of independents who are likely to be ideologically moderate and thus include the median voter (Hypothesis 2). Alternatively, other scholars (e.g., \citealt{kastellec2015polarizing}) expect that senators may respond more to co-partisans who represent both a core base of supporters and the electoral for primary elections (Hypothesis 3). Finally, we note that these expectations may be conditional on electoral context. Senators whose parties are electorally strong in a state may be more likely to respond to copartisan constituents' opinions, as electoral competition is more likely to come in a primary and the median voter is more likely to be a member of their own party. Conversely, those whose state parties are electorally weak may be better off responding to the preferences of independents (Hypothesis 4).

\subsection{Multivariate MRP Setup}

To explore the role of partisan public opinion on senators' votes, we perform multivariate MRP with three questions: party identification, opinion on the Farm Bill, and opinion on the assault weapons ban. We do so using data from the 2014 CES that asks voters for their opinion on both roll call votes; Appendix~\ref{app:responsiveness_bills_surveys} contains the exact question wording. To estimate opinion, we use the same model as in Section~\ref{sec:twostep_validation}, but also include a demographic predictor for income, as this seems to be highly predictive of individual-level responses to the question on the Farm Bill, perhaps unsurprisingly given the CES question connects it to the food stamp program.\footnote{We use household income and create five categories, following \cite{ghitza2013mrp}: under \$20k, \$20k-\$40k, \$40k-\$80k, \$80k-\$150k, and \$150k+.} This model contains 24 random effects and is estimated using around 50,000 respondents.

\begin{figure}[htp]
    \centering
    \caption{State Opinion Distribution by Party}
    \vspace{-0.5em}
    \includegraphics[width=0.85\textwidth]{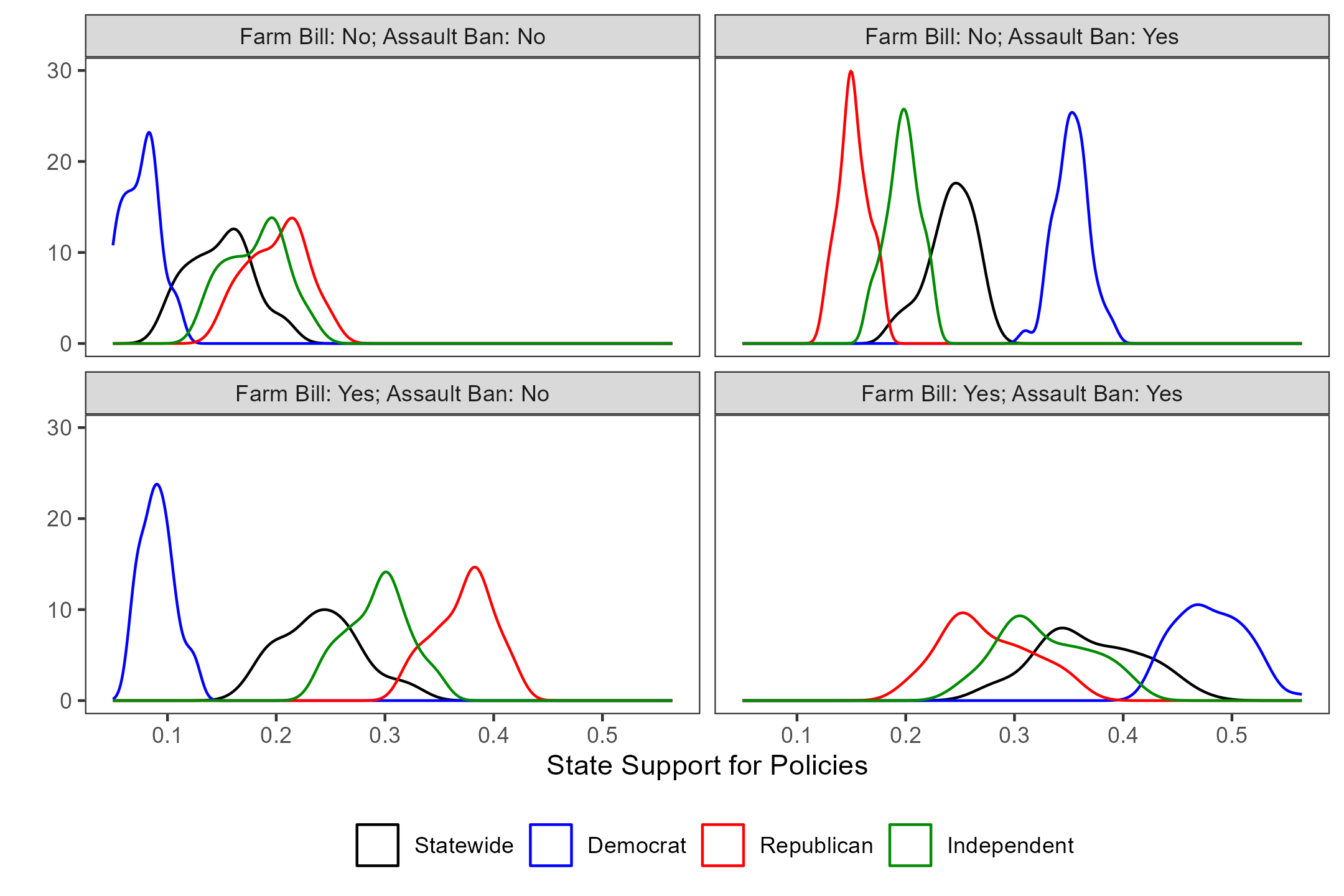}
    \label{fig:empiricdistribution}
    \caption*{\footnotesize \emph{Note:} The figure shows the distribution of opinion estimates across all 50 states for each group.}
\end{figure}

There are many quantities of interest that can be obtained from \texttt{mvMRP}. We visualize two of them here, and Appendix~\ref{app:responsivness_qoi} contains a few others (e.g., a measure of within-party heterogeneity).

Figure~\ref{fig:empiricdistribution} begins by showing the distribution of public opinion estimates from our model. It reports the distribution of state-level support for each possible combination of policies overall and by party. First, we note that in no state does any combination of policies hold support among an absolute majority of the public. Only among Democrats in 15 states does a majority of any partisan subgroup support a single set of policies (support for both the assault weapons ban and the Farm Bill). Thus, senators face a challenging decision---no combination of votes can align with the preferred policy of an absolute majority of constituents, and only rarely among copartisans. Second, the figure clearly shows a large preference among Democratic partisans for the assault weapons ban---support for the ban along with either position on the Farm Bill far exceeds support for the positions against the ban. Yet, Figure~\ref{fig:empiricdistribution} shows clear variation across states in public opinion across all sub-groups on at least some of the options under consideration.

\begin{figure}[htp]
    \centering
    \caption{Support for Cross-Pressured Positions}
    \vspace{-0.5em}
    \includegraphics[width=\textwidth]{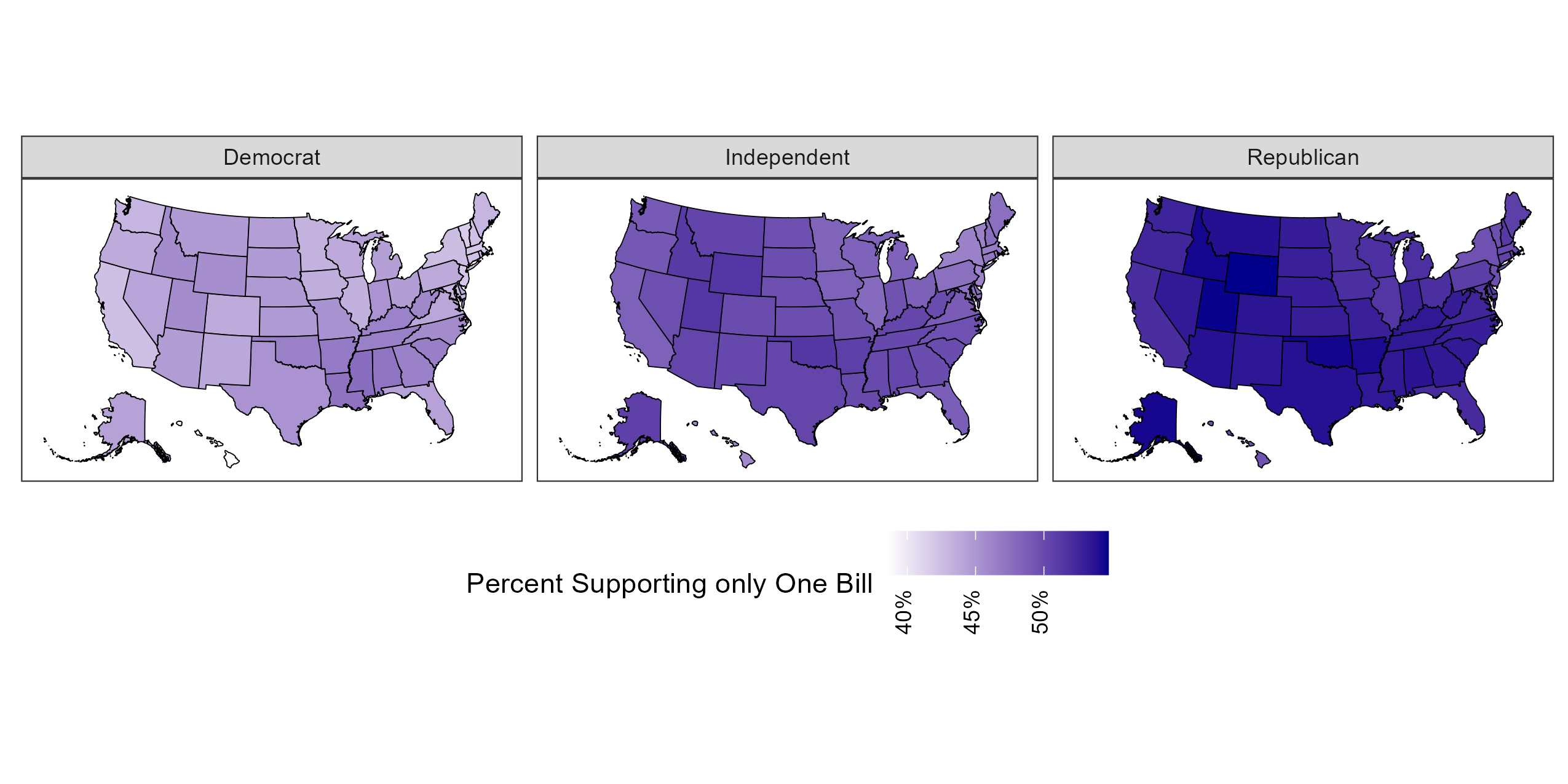}
    \label{fig:offdiagnoals}
    \caption*{\footnotesize \emph{Note:} The figure shows the share of each state and party that prefer a combination of policies with only one ``yes'' vote.}
\end{figure}

Another approach to visualizing these results is to map the interesting cases where the public is cross-pressured---that is, they support one bill but not the other. Figure~\ref{fig:offdiagnoals} shows that Democrats are less likely to be cross-pressured on these two policies; Republicans are more likely to support only one of the two bills at issue. Even among copartisan groups, there is interesting geographic variation.

\subsection{Empirical Analysis}

\begin{figure}[htbp]
    \centering
    \caption{Mass Support for Senators' Votes}
    \hspace*{-0.05\linewidth}
    \includegraphics[width=\textwidth]{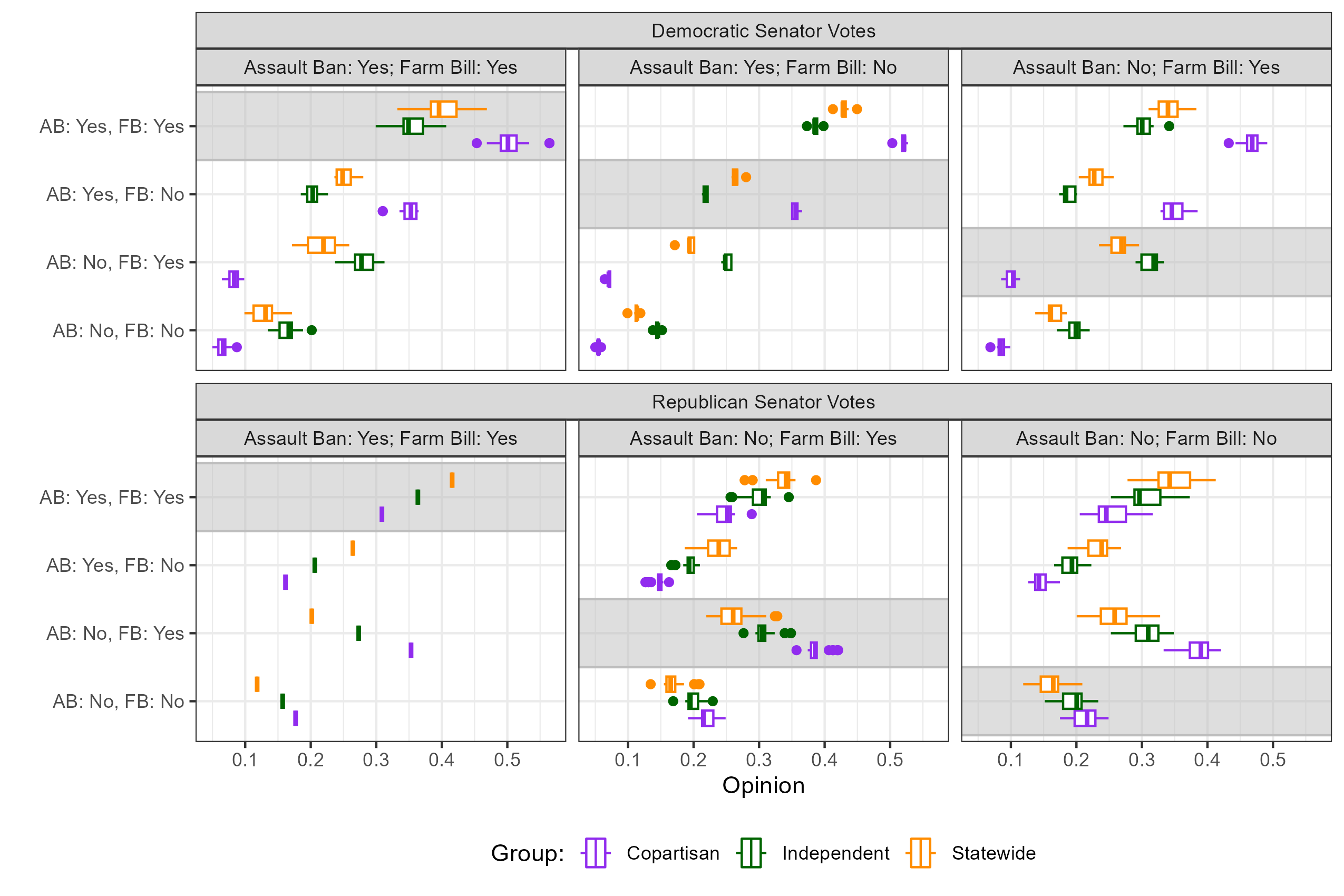}
    \label{fig:empiricrollcallopinion}
    \caption*{\footnotesize \emph{Note}: Each panel represents a party and vote choice that was observed; for example, the upper left panel contains opinion from the states of all Democratic senators who voted yes on both proposals. The opinion estimates corresponding to the actual choice of the senator are denoted with a gray box.}
\end{figure}

We now consider the relationship between opinion and legislator votes. Figure~\ref{fig:empiricrollcallopinion} begins to tackle this question by showing the distribution of opinion based on the senators' votes. Each panel reports the state opinion estimates for all states where senators cast a particular pattern of votes (e.g., the top-left reports opinion for all Democratic senators who voted against the assault weapons ban and in favor of the Farm Bill). We show opinion estimates from three possibly relevant groups given our earlier hypothesis: statewide opinion (in orange), copartisan opinion (in purple), and opinion among independents (in green).

Examining the figure shows that some senators' votes are well-explained by opinion. For example, Democrats supporting both bills generally sided with the modal citizen and modal Democrat in their states. Republicans voting against the assault weapons ban and for the Farm Bill similarly tended to agree with their modal copartisan constituent. However, there are some votes in Figure~\ref{fig:empiricrollcallopinion} that appear inconsistent with elite responsiveness to opinion. We see that, for example, Democrats who voted against the assault weapons ban and for the Farm Bill chose a position that was quite popular among independent voters---although it was extremely \emph{unpopular} among their Democratic copartisan voters.

While fully exploring this phenomenon is outside of the scope of this paper, Table~\ref{tab:empiricalregressions} reports the results of regressions that more carefully examine the hypotheses described above; Appendix~\ref{app:responsiveness_descriptives} contains descriptive statistics for the variables in these regressions. Extending the tradition of measuring dyadic representation by predicting a senator's (binary) vote on a policy using public support among their constituents (e.g., \citealt{erikson1978constituency,achen1978measuring}), we estimate a conditional logistic regression where the opinion for each combination is used as a predictor.\footnote{We exclude from the analysis Independents (Sanders and King), as well as two senators (Cowan and Chiesa) who left the chamber between the votes, and their replacements.} \citet{achen1978measuring} notes that there is no natural threshold for stating that a politician has taken a ``representative'' position when multiple dimensions are considered; instead, we consider whether greater mass support for a combination of policies increases the likelihood of a senator casting the corresponding votes.

\begin{table}[tb]
    \centering
    \caption{Predicting Senator Votes}
    
\begin{tabular}{@{\extracolsep{5pt}}lccccc} 
\\[-1.8ex]\hline \\[-1.8ex] 
\\[-1.8ex] & (1) & (2) & (3) & (4) & (5)\\ 
\hline \\[-1.8ex] 
 Statewide Op. & 0.31$^{**}$ &  &  &  &  \\ 
  & (0.06) &  &  &  &  \\ 
  Independent Op. &  & 0.31$^{**}$ &  & 0.25$^{**}$ & 0.20$^{*}$ \\ 
  &  & (0.06) &  & (0.07) & (0.09) \\ 
  Independent Op. * Pres. Share &  &  &  &  & $-$0.01$^{**}$ \\ 
  &  &  &  &  & (0.004) \\ 
  Copartisan Op. &  &  & 0.07$^{**}$ & 0.05$^{**}$ & 0.06$^{**}$ \\ 
  &  &  & (0.01) & (0.02) & (0.02) \\ 
  Copartisan Op. * Pres. Share &  &  &  &  & 0.01$^{**}$ \\ 
  &  &  &  &  & (0.003) \\ 
 N & 96 & 96 & 96 & 96 & 96 \\ 
R$^{2}$ & 0.23 & 0.20 & 0.19 & 0.26 & 0.33 \\ 
Log Likelihood & $-$93.50 & $-$97.54 & $-$98.91 & $-$89.53 & $-$80.92 \\ 
\hline \\[-1.8ex] 
\multicolumn{6}{l}{} \\ 
\end{tabular} 

    \label{tab:empiricalregressions}
    \caption*{\footnotesize \emph{Note}: $^{*}$: $p < 0.05$; $^{**}$: $p < 0.01$. All models are conditional logistic regressions; the intercepts are not shown. All variables are measured on a scale of 0 to 100. For interpretability, presidential share is centered at 50\%.}
\end{table}

The first column of Table~\ref{tab:empiricalregressions} reports the relationship between statewide opinion and legislators' votes. Consistent with the expectations from the literature and our first hypothesis, we find that greater support from the general public for a combination of policies does predict senators' likelihood of supporting it. Columns 2-4 present results for our next two hypotheses. We find that, as expected, senators are more likley to cast a combination of votes if it is popular among independents and copartisans they represent. Notably, we find in Model 4 that independent opinion is a stronger predictor than copartisan opinion.

Finally, we test Hypothesis 4---that the role of independent and copartisan opinion is conditional on the relative strength of senators' parties in their states. This may be especially useful to explain some of the more ``unusual'' decisions of senators, discussed above. To do so, we interact independent and copartisan opinion with the vote share the senator's party received in the 2012 presidential election. We use presidential vote share (rather than senators' own vote share) because it sidesteps the timing of Senate elections and focuses on relative partisan strength, rather than the dynamics of a specific campaign. It shows that in states where senators' parties are stronger in the electorate, copartisan opinion is more important for predicting their roll-call votes. We see the expected opposite story for independent opinion---senators in states with weaker presidential electoral performance rely more on independent opinion. Appendix~\ref{app:sec_responsiveness_altspecs} shows a similar story when the moderating variable is the share of copartisans as estimated from the \texttt{mvMRP} model.

\begin{figure}[tbp]
    \centering
    \caption{Moderation of Effect of Opinion}
    \includegraphics[width=0.7\textwidth]{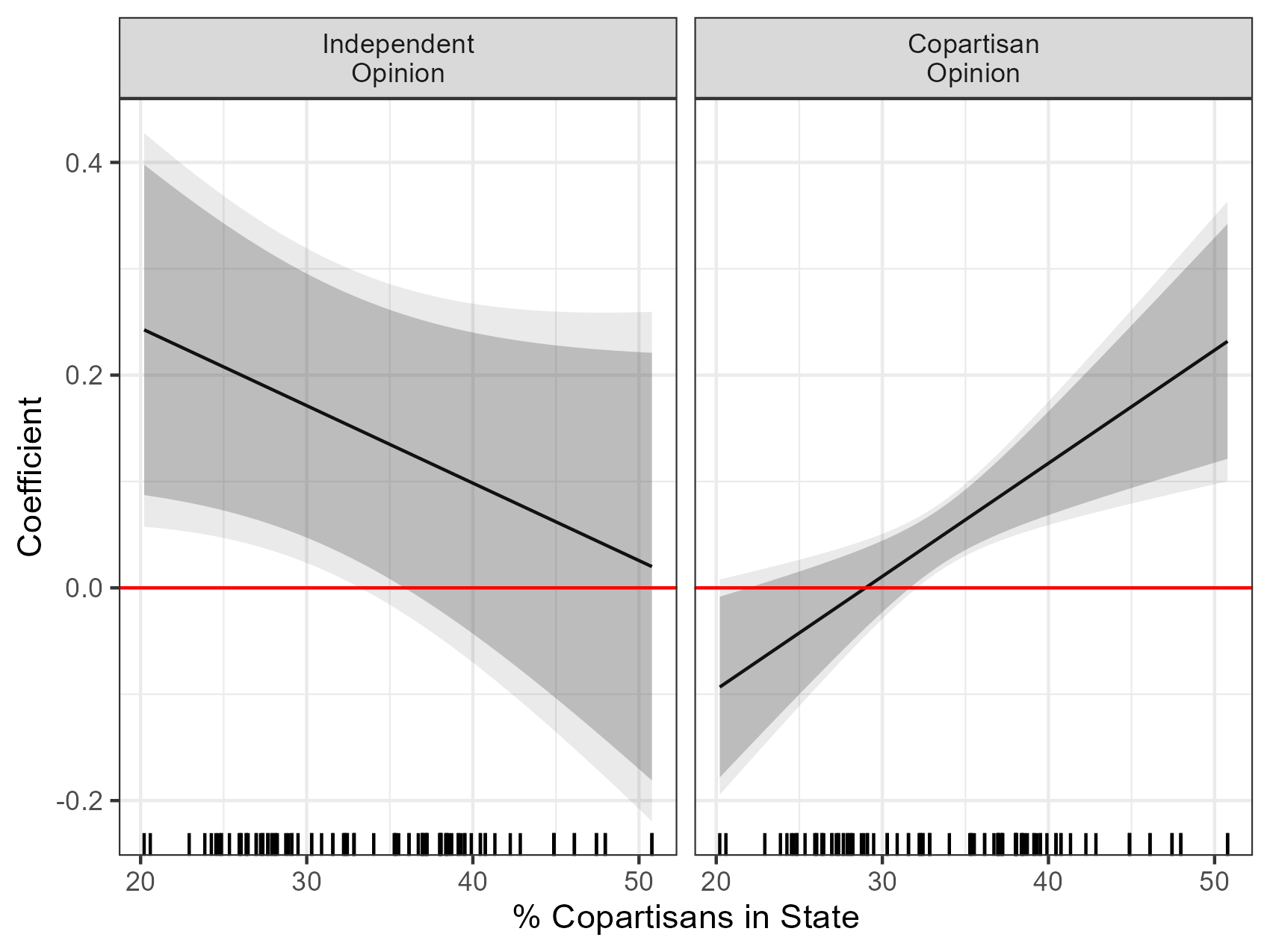}
    \label{fig:empiricslopes}
    \caption*{\footnotesize \emph{Note}: This figure shows the coefficient on each variable at different levels of presidential vote share, i.e. the main effect plus the interaction term times the moderator value. 90 and 95\% confidence intervals are shown. The distribution of the observed values of the moderator are shown in the rug plot.}
\end{figure}

To illustrate this more clearly, Figure~\ref{fig:empiricslopes} shows the coefficient on independent opinion and copartisan opinion at different levels of the presidential vote share moderator. For low levels of partisan support in the presential election, the coefficient on independent opinion is statistically significant but the coefficient on copartisan opinion is not. In cases with high same-party presidential vote, the opposite story holds: the effect of copartisan opinion is statistically significant but the effect of independent opinion is not.
 
Thus, our initial exploration of multidimensional responsiveness suggests that, similar to the typical single-issue story, the joint distribution of public opinion is correlated with the voting decisions of senators. We also find evidence that suggests that the group whose preferences matter most depends on the electoral circumstances of a senator's home state. Not all elite behavior can be well-explained by these variables, but multivariate MRP provides the ability to estimate the public opinion component of any story of legislative behavior that seeks to balance voting decisions across multiple policy areas.

\section{Conclusion}

In this paper, we proposed multivariate MRP, a method for estimating the joint distributions of public opinion on multiple survey questions at subnational geographies in a single model. We do so by reformulating the multidimensional choice (e.g., multiple policy areas, or policy areas and a characteristic such as party identification) into a combined question that can be modelled using a multinomial or conditional logistic regression and then post-stratified. To estimate this model efficiently, we employ a new method for variational inference that represents the problem as a Poisson regression with individual-specific fixed effects. We showed that even with many random effects, this can be estimated easily.

\texttt{mvMRP} produces estimates that are similar in quality to existing ``two-stage'' methods for estimating opinion by party in the United States. We also show that these estimates can be further improved by adding covariates that vary across options under consideration.

There are many potential uses for multivariate MRP. For example, existing scholarship on political representation is limited by the inability to measure how constituent public opinion is distributed across several issues at once. Our initial exploration of multidimensional representation in Section~\ref{sec:application} suggests that senators are responsive to the opinion of copartisans and independents across two policy dimensions, but that they are sensitive to the relative size of partisan groups in their states. Likewise, \texttt{mvMRP} allows scholars to understand how opinion varies across subgroups of the public that are not captured by census variables, such as by levels of political interest and among voters vs. non-voters, potentially opening up new avenues for future research.

\bibliography{hier_bib,rep_bib}

@article{goplerud2022mavb,
	title = {Fast and Accurate Estimation of Non-Nested Binomial Hierarchical Models Using Variational Inference},
	author = {Goplerud, Max},
	journal = {Bayesian Analysis},
	volume = {17},
	number = {2},
	year = {2022},
	pages = {623--650}
}

@book{gelman2006multi,
	title = {Data Analysis Using Regression and Multilevel/Hierarchical Models},
	author = {Gelman, Andrew and Hill, Jennifer},
	publisher = {Cambridge University Press},
	address = {New York},
	year = {2006}
}

@article{blei2017vi,
	author = {Blei, David M. and Kucukelbir, Alp and McAuliffe, Jon D.},
	title = {Variational Inference: A Review for Statisticians},
	journal = {Journal of the American Statistical Association},
	volume = {112},
	number = {518}, 
	year = {2017},
	pages = {859--877}	
}

@article{grimmer2011vi,
	title = {An Introduction to Bayesian Inference via Variational Approximations},
	author = {Grimmer, Justin},
	volume = {19},
	number = {1},
	year = {2011},
	journal = {Political Analysis},	
	pages = {32--47}
}

@article{bates2015lmer,
	author = {Bates, Douglas and M{ä}chler, Martin and Bolker, Ben and Walker, Steve},
	title = {Fitting Linear Mixed-Effects Models Using lme4},
	journal = {Journal of Statistical Software},
	volume = {67},
	number = {1},
	year = {2015},
	pages = {1--48}
}

@article{zhao2006mlm,
	author = {Zhao, Yihua and Staudenmayer, John and Coull, Brent A. and Wand, Matt P.},
	title = {General Design {Bayesian} Generalized Linear Mixed Models},
	journal = {Statistical Science},
	pages = {35--51},
	year = {2006},
	volume = {21},
	number = {1}
}

@article{carpenter2017stan,
	title = {Stan: A Probabilistic Programming Language},
	author = {Carpenter, Bob and Gelman, Andrew and Hoffman, Matthew D. and Lee, Daniel and Goodrich, Ben and Betancourt, Michael and Brubaker, Marcus and Guo, Jiqiang and Li, Peter and Riddell, Allen},
	journal = {Journal of Statistical Software},
	year = {2017},
	volume = {76},
	number = {1},
	pages = {1--32}
}

@article{burkner2017brms,
	title = {brms: An R Package for Bayesian Multilevel Models Using Stan},
	author = {B{ü}rkner, Paul-Christian},
	journal = {Journal of Statistical Software},
	year = {2017},
	volume = {80},
	number = {1},
	pages = {1--28}
}

@article{greeneRLS1991,
	year  = {1991},
	volume = {73},
	number = {3},
	pages = {563--567},
	author = {William H. Greene and Terry G. Seaks},
	title = {The Restricted Least Squares Estimator: A Pedagogical Note},
	journal = {The Review of Economics and Statistics}
}

@book{lawson1974linear,
	title = {Solving Least Squares Problems},
	author = {Lawson, Charles L. and Hanson, Richard J.},
	publisher = {Prentice-Hall},
	address = {Englewood Cliffs, NJ},
	year = {1974}
}

@article{knill2014pseudo,
	year = {2014},
	volume = {459},
	pages = {522--547},
	author = {Oliver Knill},
	title = {Cauchy{\textendash}Binet for Pseudo-Determinants},
	journal = {Linear Algebra and its Applications}
}

@article{ghitza2013mrp,
	year = {2013},
	volume = {57},
	number = {3},
	pages = {762--776},
	author = {Yair Ghitza and Andrew Gelman},
	title = {Deep Interactions with {MRP}: Election Turnout and Voting Patterns Among Small Electoral Subgroups},
	journal = {American Journal of Political Science}
}

@article{park2004mrp,
	year = {2004},
	volume = {12},
	number = {4},
	pages = {375--385},
	author = {David K. Park and Andrew Gelman and Joseph Bafumi},
	title = {Bayesian Multilevel Estimation with Poststratification: State-Level Estimates from National Polls},
	journal = {Political Analysis}
}

@article{lax2009estimation,
	year = {2009},
	volume = {53},
	number = {1},
	pages = {107--121},
	author = {Jeffrey R. Lax and Justin H. Phillips},
	title = {How Should We Estimate Public Opinion in The States?},
	journal = {American Journal of Political Science}
}

@article{imai2016ideal,
	year = {2016},
	volume = {110},
	number = {4},
	pages = {631--656},
	author = {Kosuke Imai and James Lo and Jonathan Olmsted},
	title = {Fast Estimation of Ideal Points with Massive Data},
	journal = {American Political Science Review}
}

@article{ornstein2019stacked,
	year = {2020},
	pages = {293--301},
	volume = {28},
	number = {2},
	author = {Joseph T. Ornstein},
	title = {Stacked Regression and Poststratification},
	journal = {Political Analysis}
}

@book{bishop2006ml,
	title = {Pattern Recognition and Machine Learning},
	author = {Bishop, Christopher},
	year = {2006},
	publisher = {Springer},
	address = {New York}
}

@article{buttice2013mrp,
	year = {2013},
	volume = {21},
	number = {4},
	pages = {449--467},
	author = {Matthew K. Buttice and Benjamin Highton},
	title = {How Does Multilevel Regression and Poststratification Perform with Conventional National Surveys?},
	journal = {Political Analysis}
}

@book{agresti_categorical_2002,
	address = {Hoboken, NJ},
	title = {Categorical {Data} {Analysis}},
	publisher = {John Wiley \& Sons, Inc.},
	author = {Agresti, Alan},
	year = {2002}
}

@incollection{mcfadden_conditional_1974,
	publisher = {Academic Press},
	address = {New York},
	year = {1974},
	booktitle = {{Frontiers in Econometrics}},
	title = {Conditional {Logit} {Analysis} of {Qualitative} {Choice} {Behavior}},
	author = {McFadden, Daniel},
	editor = {Zaremmbka, Paul}
}

@article{caughey_dynamic_2015,
	title = {Dynamic Estimation of Latent Opinion Using a Hierarchical Group-Level {IRT} Model},
	volume = {23},
	pages = {197--211},
	number = {2},
	journal = {Political Analysis},
	author = {Caughey, Devin and Warshaw, Christopher},
	year = {2015}
}

@article{mcfadden2000mixed,
	author = {McFadden, Daniel and Train,Kenneth E.},
	title = {Mixed MNL Models for Discrete Response},
	journal = {Journal of Applied Econometrics},
	year = {2000},
	volume = {15},
	number = {5},
	pages = {447-470}
}

@article{warshaw2012district,
  year = {2012},
  volume = {74},
  number = {1},
  pages = {203--219},
  author = {Christopher Warshaw and Jonathan Rodden},
  title = {How Should We Measure District-Level Public Opinion on Individual Issues?},
  journal = {The Journal of Politics}
}

@article{broniecki2021auto,
	title={Improved Multilevel Regression with Post-Stratification Through Machine Learning (autoMrP)},
	author={Broniecki, Philipp and Leemann, Lucas and W{\"u}est, Reto},
	journal={The Journal of Politics},
	volume = {84},
	number = {1},
	pages = {597--601},
	year = {2022}
}

@article{goplerud2023reeval,
	title={Re-Evaluating Machine Learning for MRP Given the Comparable Performance of (Deep) Hierarchical Models},
	author = {Goplerud, Max},
	year = {2024},
        volume = {118},
        number = {1},
        pages = {529--536},
	journal = {American Political Science Review},
        doi = {10.1017/S0003055423000035}
}

@article{ameli2023singular,
  title={A Singular Woodbury and Pseudo-Determinant Matrix Identities and Application to Gaussian Process Regression},
  author={Ameli, Siavash and Shadden, Shawn C},
  journal={Applied Mathematics and Computation},
  volume={452},
  pages={128032},
  year={2023},
  publisher={Elsevier}
}

@article{wand2014fully,
  title={Fully Simplified Multivariate Normal Updates in Non-conjugate Variational Message Passing},
  author={Wand, Matt P},
  journal={Journal of Machine Learning Research},
  year={2014},
  volume = {15},
  pages = {1351--1369}
}

@article{knowles2011non,
  title={Non-conjugate Variational Message Passing for Multinomial and Binary Regression},
  author={Knowles, David and Minka, Tom},
  journal={Advances in Neural Information Processing Systems},
  volume={24},
  year={2011}
}

@article{guimaraes2003tractable,
  title={A Tractable Approach to the Firm Location Decision Problem},
  author={Guimaraes, Paulo and Figueirdo, Oct{\'a}vio and Woodward, Douglas},
  journal={Review of Economics and Statistics},
  volume={85},
  number={1},
  pages={201--204},
  year={2003}
}

@article{baker1994multinomial,
  title={The Multinomial-Poisson Transformation},
  author={Baker, Stuart G},
  journal={Journal of the Royal Statistical Society: Series D (The Statistician)},
  volume={43},
  number={4},
  pages={495--504},
  year={1994},
  publisher={Wiley Online Library}
}

@article{tan2017stochastic,
  title={Stochastic Variational Inference for Large-scale Discrete Choice Models using Adaptive Batch Sizes},
  author={Tan, Linda S. L.},
  journal={Statistics and Computing},
  volume={27},
  pages={237--257},
  year={2017},
  publisher={Springer}
}

@article{bansal2020bayesian,
  title={Bayesian Estimation of Mixed Multinomial Logit Models: Advances and Simulation-based Evaluations},
  author={Bansal, Prateek and Krueger, Rico and Bierlaire, Michel and Daziano, Ricardo A and Rashidi, Taha H},
  journal={Transportation Research Part B: Methodological},
  volume={131},
  pages={124--142},
  year={2020},
  publisher={Elsevier}
}

@book{wood2017gam,
  title={Generalized Additive Models: An Introduction with R},
  author={Wood, Simon N},
  year={2017},
  publisher={CRC press},
  address={New York}
}

@book{train2009discrete,
  title={Discrete Choice Methods with Simulation},
  author={Train, Kenneth E},
  year={2009},
  address = {New York},
  publisher={Cambridge University Press}
}

@article{braun2010discrete,
  title={Variational Inference for Large-scale Models of Discrete Choice},
  author={Braun, Michael and McAuliffe, Jon},
  journal={Journal of the American Statistical Association},
  volume={105},
  number={489},
  pages={324--335},
  year={2010},
  publisher={Taylor \& Francis}
}

@article{leemann2017extending,
  title={Extending the Use and Prediction Precision of Subnational Public Opinion Estimation},
  author={Leemann, Lucas and Wasserfallen, Fabio},
  journal={American Journal of Political Science},
  volume={61},
  number={4},
  pages={1003--1022},
  year={2017}
}

@article{taddy2015distributed,
author = {Matt Taddy},
title = {Distributed Multinomial Regression},
volume = {9},
journal = {The Annals of Applied Statistics},
number = {3},
pages = {1394--1414},
year = {2015}
}

@article{marble2025improving,
    author = {Marble, William and Clinton, Joshua},
    title = {Improving Small-area Estimates of Public Opinion by Calibrating to Known Population Quantities},
    journal = {SocArXiv},
    year = {2024}, 
    url = {doi.org/10.31235/osf.io/u3ekq}
}

@article{rifkin2004defense,
  title={In defense of one-vs-all classification},
  author={Rifkin, Ryan and Klautau, Aldebaro},
  journal={Journal of Machine Learning Research},
  volume={5},
  number={Jan},
  pages={101--141},
  year={2004}
}

@Manual{elff2025mclogit,
  title = {mclogit: Multinomial Logit Models, with or without Random Effects or Overdispersion},
  author = {Martin Elff},
  year = {2025},
  note = {R package version 0.9.10},
  url = {http://melff.github.io/mclogit/},
}

@article{croissant2020mlogit,
 title={Estimation of Random Utility Models in R: The mlogit Package},
 volume={95},
 number={11},
 journal={Journal of Statistical Software},
 author={Croissant, Yves},
 year={2020},
 pages={1--41}
}

@article{lax2019party,
    author = {Lax, Jeffrey R. and Phillips, Justin H. and Zelizer, Adam},
    title = {The Party or the Purse? Unequal Representation in the US Senate},
    journal = {American Political Science Review},
    year = {2019},
    volume = {113},
    number = {4},
    pages = {917--940}
}

@article{achen1978measuring,
    author = {Achen, Christopher H.},
    title = {Measuring Representation},
    journal = {American Journal of Political Science},
    year = {1978},
    volume = {22},
    number = {3},
    pages = {457--510}
}

@article{erikson1978constituency,
    author = {Erikson, Robert S.},
    title = {Constituency Opinion and Congressional Behavior: A Reexamination of the Miller-Stokes Representation Data},
    journal = {American Journal of Political Science},
    year = {1978},
    volume = {22},
    number = {3},
    pages = {511--535}
}

@article{page1983effects,
    author = {Page, Benjamin I. and Shapiro, Robert Y.},
    title = {Effects of Public Opinion on Policy},
    journal = {American Political Science Review},
    year = {1983},
    volume = {77},
    number = {1},
    pages = {175--190}
}

@article{lax2012democratic,
    author = {Lax, Jeffrey R. and Phillips, Justin H.},
    title = {The Democratic Deficit in the States},
    journal = {American Journal of Political Science},
    year = {2012},
    volume = {56},
    number = {1},
    pages = {148--166}
}

@article{broockman2016approaches,
    author = {Broockman, David E.},
    title = {Approaches to Studying Policy Representation},
    journal = {Legislative Studies Quarterly},
    year = {2016},
    volume = {41},
    number = {1},
    pages = {181--215}
}

@book{dahl1956preface,
    title = {A Preface to Democratic Theory},
    publisher = {University of Chicago Press},
    address = {Chicago},
    author = {Dahl, Robert A.},
    year = {1956}
}

@book{arnold1990logic,
    title = {The Logic of Congressional Action},
    publisher = {Yale University Press},
    address = {New Haven, CT},
    author = {Arnold, R. Douglas},
    year = {1990}
}

@article{kastellec2015polarizing,
  title={Polarizing the Electoral Connection: Partisan Representation in Supreme Court Confirmation Politics},
  author={Kastellec, Jonathan P and Lax, Jeffrey R and Malecki, Michael and Phillips, Justin H},
  journal={The Journal of Politics},
  volume={77},
  number={3},
  pages={787--804},
  year={2015}
}

@misc{lopezmartin2022multilevel,
  title = {Multilevel Regression and Poststratification Case Studies},
  author = {Lopez-Martin, Juan and Phillips, Justin H. and Gelman, Andrew},
  year = {2022},
  howpublished = {\url{https://bookdown.org/jl5522/MRP-case-studies/}},
  note = {Accessed on July 3, 2025}
}

@article{broockman2018bias,
  title = {Bias in Perceptions of Public Opinion among Political Elites},
  author = {Broockman, David E. and Skovron, Christopher},
  year = {2018},
  journal = {American Political Science Review},
  volume = {112},
  number = {3},
  pages = {542--563}
}

@article{wright1989policy,
    author = {Wright, Gerald C.},
    title = {Policy Voting in the U.S. Senate: Who Is Represented?},
    journal = {Legislative Studies Quarterly},
    year = {1989},
    volume = {14},
    number = {4},
    pages = {465--486}
}

@manual{ruggles2025ipums,
    author = {Ruggles, Steven and Flood, Sarah and Sobek, Matthew and Backman, Daniel and Cooper, Grace and Rivera Drew, Julia A. and Richards, Stephanie and Rodgers, Renae and Schroeder, Jonathan and Williams, Kari C.W.},
    title = {IPUMS USA: Version 16.0 [dataset]},
    year = {2025},
    url = {https://doi.org/10.18128/D010.V16.0},
    address = {Minneapolis, MN},
    organization = {IPUMS}
}

@manual{dagonel2023cces,
    author = {Dagonel, Angelo},
    title = {Cumulative CES Policy Preferences [dataset]},
    year = {2021},
    url = {https://doi.org/10.7910/DVN/OSXDQO},
    organization = {Harvard Dataverse},
}
\bibliographystyle{apsr}

\appendix
\singlespacing

\section{Variational Inference with Linear Constraints}
\label{app:VI_FE}

This appendix derives the optimal variational distribution for a Gaussian likelihood with linear equality constraints on parameter $\bm{\gamma}$ such that $\bm{L}\bm{\gamma} = \bm{0}$. It further shows that if $\bm{\gamma}$ represents (high-dimensional) fixed effects, then inference can be performed extremely efficiently---in linear computational cost and memory. Thus, this establishes that representing multinomial logistic regression as Poisson regression with high-dimensional fixed effects and performing CAVI assuming independence between the fixed effects and other parameters is a computationally feasible methodology even for large problems.

Result~\ref{result:fe}, stated below, first considers the case of arbitrary equality constraints.

\begin{result}(Variational Inference with Linear Equality Constraints)
	\label{result:fe}
	Define $\bm{X}$ as an $N \times p$ matrix with rank $p$. Define $\bm{L}$ as a $C \times p$ matrix with rank $C$. Consider the coordinate ascent variational inference (CAVI) update of $q(\bm{\gamma})$ where the variational distribution of all other parameters $\bm{\theta}$ is $q(\bm{\theta})$ and $q(\bm{\theta},\bm{\gamma}) = q(\bm{\theta})q(\bm{\gamma})$. Assume the expectation of the log-posterior with respect to $q(\bm{\theta})$ is as follows and $\bm{L}\bm{\gamma} = \bm{0}$.
	
	\begin{equation*}
	E_{q(\bm{\theta})}\left[\ln p(\bm{y}, \bm{\theta}, \bm{\gamma})\right] \propto -\frac{1}{2} || \bm{v} - \bm{X} \bm{\gamma}||_2^2 \quad \mathrm{s.t.} \quad \bm{L}\bm{\gamma}=\bm{0}    
	\end{equation*}
	
	Then, the optimal approximating distribution on $\bm{\gamma}$ is a \emph{singular} multivariate Gaussian with rank $p-C$ with mean $\bm{\mu}_{\bm{\gamma}}$ and variance $\bm{\Lambda}_{\bm{\gamma}}$ as shown below as well as the log pseudo-determinant (the sum of the log of the non-zero singular values) of $\bm{\Lambda}_{\bm{\gamma}}$, i.e. $\ln|\bm{\Lambda}_{\bm{\gamma}}|_+$,
	
	\begin{align*}
	\hat{\bm{\gamma}}_{OLS} &= \left(\bm{X}^T\bm{X}\right)^{-1} \bm{X}^T\bm{v}; ~\bm{P}_{\bm{L},\bm{X}} = \bm{L}^T\left(\bm{L}\left[\bm{X}^T\bm{X}\right]^{-1}\bm{L}^T\right)^{-1}\bm{L} \\
	\bm{\mu}_{\bm{\gamma}} &= \hat{\bm{\gamma}}_{OLS} - \left(\bm{X}^T\bm{X}\right)^{-1} \bm{P}_{\bm{L},\bm{X}}~\hat{\bm{\gamma}}_{OLS}; \quad \bm{\Lambda}_{\bm{\gamma}} = \left(\bm{X}^T\bm{X}\right)^{-1} - \left(\bm{X}^T\bm{X}\right)^{-1} \bm{P}_{\bm{L},\bm{X}}\left(\bm{X}^T\bm{X}\right)^{-1} \\
	\ln|\bm{\Lambda}_{\bm{\gamma}}|_{+} &= \ln|\bm{L}\bm{L}^T| - \ln|\bm{X}^T\bm{X}| - \ln|\bm{L}\left(\bm{X}^T\bm{X}\right)^{-1}\bm{L}^T|
	\end{align*}
	
	Further, the ELBO takes the following form:
	\begin{align*}
	\mathrm{ELBO} = \mathbb{E}_{q(\bm{\theta})q(\bm{\gamma})}\left[\ln p(\bm{y}, \bm{\theta}, \bm{\gamma})\right] - \mathbb{E}_{q(\bm{\theta})}[\ln q(\bm{\theta})] + \frac{1}{2} \ln|\bm{\Lambda}_{\bm{\gamma}}|_+
	\end{align*}
\end{result}

\begin{proof}
	Define $\bm{M}$ as any orthogonal basis of the nullspace of $\bm{L}$, i.e. $\bm{L}\bm{M}\bm{a} = \bm{0}$ for any vector $\bm{a}$ and $\bm{M}^T\bm{M} = \bm{I}$. Standard results on constrained least squares (e.g., \citealt{lawson1974linear}) imply that one can re-parameterize the problem to perform inference on a $p-C$ unconstrained parameter vector $\bm{\gamma}^*$ where $\bm{\gamma} = \bm{M}\bm{\gamma}^*$. In this case, the expectation of the log-posterior over $q(\bm{\theta})$ becomes
	\begin{equation*}
	E_{q(\bm{\theta})}\left[\ln p(\bm{y}, \bm{\theta}, \bm{\gamma}^*)\right] \propto -\frac{1}{2} || \bm{v} - \bm{X} \bm{M} \bm{\gamma}^*||_2^2.
	\end{equation*}
	
	Standard results (e.g., \citealt{bishop2006ml}) give the optimal variational approximation as
	
	$$q(\bm{\gamma}^*) \sim N\left(\left[\bm{M}^T\bm{X}^T\bm{X}\bm{M}\right]^{-1} \bm{M}^T\bm{X}^T\bm{v}, \left[\bm{M}^T\bm{X}^T\bm{X}\bm{M}\right]^{-1} \right).$$
	
	If we return to the parameterization on $\bm{\gamma}$, the optimal variational approximation is 
	
	$$q(\bm{\gamma}) \sim N\left(\bm{M}\left[\bm{M}^T\bm{X}^T\bm{X}\bm{M}\right]^{-1} \bm{M}^T\bm{X}^T\bm{v}, \bm{M}\left[\bm{M}^T\bm{X}^T\bm{X}\bm{M}\right]^{-1}\bm{M}^T \right).$$
	
	Given that $\bm{X}^T\bm{X}$ is invertible, the following identity holds
	\begin{align*}
	&\bm{M}\left[\bm{M}^T\bm{X}^T\bm{X}\bm{M}\right]^{-1}\bm{M}^T = \left(\bm{X}^T\bm{X}\right)^{-1}\left[\bm{X}^T\bm{X} - \bm{P}_{\bm{L},\bm{X}} \right]\left(\bm{X}^T\bm{X}\right)^{-1} \\
	&\bm{P}_{\bm{L},\bm{X}} = \bm{L}^T\left(\bm{L}\left[\bm{X}^T\bm{X}\right]\bm{L}\right)^{-1}\bm{L}.
	\end{align*}
	Thus, $q(\bm{\gamma})$ can be re-arranged to as follows:
	\begin{align*}
	q(\bm{\gamma}) &\sim N\left(\bm{\mu}_{\bm{\gamma}}, \bm{\Lambda}_{\bm{\gamma}}\right) \\
	\bm{\mu}_{\bm{\gamma}} &= \left(\bm{X}^T\bm{X}\right)^{-1}\left[\bm{X}^T\bm{X} - \bm{P}_{\bm{L},\bm{X}}\right] \hat{\bm{\gamma}}_{OLS}; \quad \hat{\bm{\gamma}}_{OLS} = \left(\bm{X}^T\bm{X}\right)^{-1}\bm{X}^T\bm{\nu} \\
	\bm{\Lambda}_{\bm{\gamma}} &= \left(\bm{X}^T\bm{X}\right)^{-1}\left[\bm{X}^T\bm{X} - \bm{P}_{\bm{L},\bm{X}} \right]\left(\bm{X}^T\bm{X}\right)^{-1}
	\end{align*}
	\cite{ameli2023singular} provide results on evaluating pseudo-determinants; their results can be used to show that
	
	$$\ln|\bm{\Lambda}_{\bm{\gamma}}|_+ = \ln|\bm{L}\bm{L}^T| - \ln|\bm{X}^T\bm{X}| - \ln|\bm{L}\left(\bm{X}^T\bm{X}\right)^{-1} \bm{L}^T|.$$
	
	This establishes the first part of the result. For evaluating the ELBO, consider if inference were performed on the unconstrained $\bm{\gamma}^*$. In this case,
	$$\mathrm{ELBO} = \mathbb{E}_{q(\bm{\theta})q(\bm{\gamma}^*)}\left[\ln p(\bm{y}, \bm{\theta}, \bm{\gamma}^*)\right] - \mathbb{E}_{q(\bm{\theta})}[\ln q(\bm{\theta})] + \frac{1}{2} \ln\left|\left[\bm{M}^T\bm{X}^T\bm{X}\bm{M}\right]^{-1}\right|.$$
	
	Note that if we re-parameterize and use $q(\bm{\gamma})$ instead of $q(\bm{\gamma}^*)$, the first term remains unchanged by definition. The final term, the entropy of $q(\bm{\gamma}^*)$, can be evaluated as follows. \cite{knill2014pseudo} establishes the following identity of pseudo-determinants: $|\bm{F}^T\bm{G}|_+ = |\bm{F}\bm{G}^T|_+$. Noting this and that $\bm{M}^T\bm{M} = \bm{I}$,
	$$\left|\left[\bm{M}^T\bm{X}^T\bm{X}\bm{M}\right]^{-1}\right| = \left|\bm{\Lambda}_{\bm{\gamma}}\right|_+.$$
	
	To establish the remaining claim, we re-parameterize the ELBO in terms $\bm{\gamma}$, 
	$$\mathrm{ELBO} = \mathbb{E}_{q(\bm{\theta})q(\bm{\gamma})}\left[\ln p(\bm{y}, \bm{\theta}, \bm{\gamma})\right] - \mathbb{E}_{q(\bm{\theta})}[\ln q(\bm{\theta})] + \frac{1}{2} \ln\left|\bm{\Lambda}_{\bm{\gamma}}\right|_+.$$
\end{proof}

In interpreting Result~\ref{result:fe}, we thus note that CAVI can be performed on $\bm{\gamma}$, and is equivalent to performing CAVI on the unconstrained $\bm{\gamma}^*$, but it does not require explicitly forming or using a basis $\bm{M}$. We note that the optimal variational distribution $q(\bm{\gamma})$ also exactly coincides with the distribution of an OLS estimator with linear constraints (e.g., \citealt{greeneRLS1991}).

However, in practice, Result~\ref{result:fe} is not computationally feasible for large-scale problems. Consider the case of fixed effects where $\bm{X}$ is high-dimensional, i.e. $p$ is large, and $\bm{1}^T\bm{\gamma} = 0$. Note that $q(\bm{\gamma})$ is a \textbf{dense} $p \times p$ (singular) multivariate Gaussian and thus extremely computationally expensive to store and manipulate. Result~\ref{result:simple_fe} shows that in the case of fixed effects, this high cost can be avoided.

\begin{result}{(Simplified Inference for Classical Fixed Effects)}
	\label{result:simple_fe}
	Assume that $\bm{\gamma}$ has a variational distribution as defined from Result~\ref{result:fe}. Further assume that each row $\bm{x}_i$ of $\bm{X}$ is ``one-hot'', i.e. has one element with value 1 and all others with value 0. Define $\bm{d}$ as a vector corresponding to the diagonal elements of $\bm{X}^T\bm{X}$  and that $\bm{L} = \bm{1}^T$. Define $\bm{u}$ as a vector such that $u_j = 1/d_j$. $\bm{a} \odot \bm{b}$ denotes the Hadamard (elementwise) product. Then, the mean, the diagonal of $\bm{\Lambda}_{\bm{\gamma}}$ and log pseudo-determinant of $q(\bm{\gamma})$ are shown below. These can all be evaluated and stored at cost $\mathcal{O}(p)$.
	$$\hat{\bm{\gamma}}_{\mathrm{OLS}} = \bm{u} \odot \left(\bm{X}^T\bm{v}\right)$$
	$$\bm{\mu}_\gamma = \hat{\bm{\gamma}}_{OLS} - \bm{u} \cdot \frac{\left[\hat{\bm{\gamma}}_{OLS}\right]^T \bm{1}}{\bm{u}^T\bm{1}}$$
	$$\mathrm{diagonal}(\bm{\Lambda}_{\bm{\gamma}}) =  \bm{u} - \left(\bm{u} \odot \bm{u}\right) \cdot \frac{1}{\bm{u}^T\bm{1}} $$
	$$\ln\mathrm{det}^*(\bm{\Lambda}_{\bm{\gamma}}) =  \ln(\bm{u})^T\bm{1}  - \ln|\bm{u}^T\bm{1}|$$
	
	Further, $E_{q(\bm{\gamma})}[\bm{x}_i^T\bm{\gamma}]$ and $\mathrm{Var}_{q(\bm{\gamma})}[\bm{x}_i^T \bm{\gamma}]$ can be evaluated at $\mathcal{O}(1)$ for any row $\bm{x}_i$ of $\bm{X}$.
\end{result}
\begin{proof}
	The simplification of the moments and log-determinant of $q(\bm{\gamma})$ follows directly with simple algebraic re-arrangement from Result~\ref{result:fe}, noting $\bm{X}^T\bm{X}$ is a diagonal matrix by definition. For the second claim, as $\bm{x}_i$ is a one-hot vector, define index $j[i]$ as the position of the single non-zero element for row $i$. Then, note that $\bm{x}_i^T\bm{\gamma} = \bm{\gamma}_{j[i]}$, i.e. it picks out only the $j$-th element of $\bm{\gamma}$. Thus, its mean and variance can be directly read off of the simplified moments above. 
\end{proof}

The key implication of Result~\ref{result:simple_fe} is that, in the case of fixed effects---i.e. a sparse $\bm{X}$ with each row being one-hot---inference can be performed at a cost and memory requirement that scales linearly in the number of fixed effects $p$. Thus, even for huge problems, CAVI can be implemented cheaply as the only terms in the ELBO that depend on $\bm{\gamma}$ are the log-determinants and functions of the mean and variance of $\bm{x}_i^T\bm{\gamma}$. 

Interestingly, even if there are multiple high-dimensional fixed effects, this variational model can be easily implemented assuming independence between the fixed effects. Exploring this for other high-dimensional fixed effects problems is an interesting area for future research.

\section{Variational Inference for Poisson Likelihood}
\label{app:VI_poisson}

This appendix derives the ELBO and coordinate ascent variational inference (CAVI) algorithm for estimating the optimal variational distribution. The exposition follows the ``general design'' notation from \cite{zhao2006mlm}. Equation~\ref{eq:model_gendesign} replicates the generative model in \cite{goplerud2022mavb}, adapting only the likelihood to be Poisson. As that paper notes,

\begin{quote}
	``[this model] uses conditionally conjugate Inverse-Wishart priors on the variance components $\bm{\Sigma}_j$, a flat prior on the fixed effects $\bm{\beta}$, and the conventional normal prior on the random effects. $\bm{z}^b_{i,j}$ represents the covariates for individual $i$ for random effect $j$, e.g. $\bm{z}^b_{i,j} = 1$ for a random intercept. Each random effect $j \in \{1, \cdots, J\}$ has $g_j$ levels ... and a dimensionality of $d_j$ (e.g., $d_j = 1$ for a random intercept). \end{quote}

\begin{align}
\label{eq:model_gendesign}
\begin{split}
&y_i | \bm{\beta}, \bm{\alpha} \sim \mathrm{Poisson}(\lambda_i), \quad \lambda_i = \exp(\psi_i), \quad \psi_i = \bm{x}_i^T\bm{\beta} + \bm{z}_i^T\bm{\alpha} \\
&\bm{\alpha}_{j} | \bm{\Sigma}_j \sim N\left(\bm{0}, \bm{I}_{g_j} \otimes \bm{\Sigma}_j\right), \quad \bm{\Sigma}_j \sim \mathrm{IW}(\nu_j, \bm{\Phi}_j), \quad p(\bm{\beta}) \propto 1  \\
& \bm{z}_{i,j} = \bm{m}_{i,j} \otimes \bm{z}^b_{i,j}, \quad \bm{\alpha}^T = [\bm{\alpha}^T_1, \cdots, \bm{\alpha}^T_J], \quad \bm{z}_i^T = [\bm{z}_{i,1}^T, \cdots, \bm{z}_{i,J}^T]
\end{split}
\end{align}

A second presentation of the model follows the notation in \cite{gelman2006multi} and separately presents each random effect.

\begin{align}
\label{eq:model_gh}
\begin{split}
y_i | \bm{\beta}, \{\{\bm{\alpha}_{j,g}\}_{g=1}^{g_j}\}_{j=1}^J &\sim \mathrm{Poisson}(\lambda_i), \quad \lambda_i = \exp(\psi_i), \quad \psi_i = \bm{x}_i^T\bm{\beta} + \sum_{j=1}^J \left[\bm{z}^b_{i,j}\right]^T \bm{\alpha}_{j, g[i]} \\
&\bm{\alpha}_{j,g} | \bm{\Sigma}_j \sim N(\bm{0}_{d_j}, \bm{\Sigma}_j), \quad \bm{\Sigma}_j \sim \mathrm{IW}(\nu_j, \bm{\Phi}_j) \quad \forall (j,g), \quad p(\bm{\beta}) \propto 1
\end{split}
\end{align}

Following existing work, we assume that the variational distributions $q(\bm{\beta}, \{\bm{\alpha}_j\}_{j=1}^J)$ factorizes into independent distributions: $q(\bm{\beta})\prod_{j=1}^J q(\bm{\alpha}_j)$. As is common for variational inference with Poisson outcomes (e.g., \citealt{wand2014fully}), we further assume that each of those variational distributions has a multivariate Gaussian form. As is standard, we assume that $q(\{\bm{\Sigma}_j\}_{j=1}^J)$ is independent from the other parameters. This is sufficient to imply that the variational distribution of each $\bm{\Sigma}_j$ is independent from all others.

With this in hand, the ELBO can be derived as shown below (e.g., \citealt{wand2014fully}). Define $\mathrm{diagonal}(\bm{W})$ to take the diagonal elements of some matrix $\bm{W}$ and $\mathrm{diag}(\bm{w})$ to create a diagonal matrix with some vector $\bm{w}$ along the diagonal. For simplicity, we only show the portion of the ELBO that involves $q(\bm{\alpha}_j)$ and $q(\bm{\beta})$ as all updates for $q(\bm{\Sigma}_j)$ are standard. We show this using the most compact notation although note that $\bm{\Lambda}_{\alpha,j}$ is (block) diagonal.

\begin{equation}
\begin{split}
&\bm{w} = \exp\left(\bm{X} \bm{\mu}_{\beta} + \frac{1}{2} \mathrm{diagonal}\left[\bm{X} \bm{\Lambda}_{\beta}\bm{X}^T\right] + \sum_{j=1}^T \bm{Z}_j\bm{\mu}_{\alpha,j} + \frac{1}{2} \mathrm{diagonal}\left[\bm{Z} \bm{\Lambda}_{\alpha,j}\bm{Z}^T\right]\right) \\ 
& [\bm{w}]_i = \exp\left(\bm{x}_i^T\bm{\mu}_\beta + \frac{1}{2} \bm{x}_i^T\bm{\Lambda}_{\beta} \bm{x}_i + \sum_{j=1}^J \bm{z}_{i,j}^T\bm{\mu}_{\alpha,j} + \frac{1}{2} \bm{z}_{i,j}^T\bm{\Lambda}_{\alpha,j} \bm{z}_{i,j} \right) \\
&\mathrm{ELBO}_{q(\bm{\theta})} =~ \cdots + \bm{y}^T\left[\bm{X} \bm{\mu}_{\bm{\beta}} + \bm{Z}\bm{\mu}_{\bm{\alpha}}\right] - \bm{1}^T\bm{w} + \\
&\frac{1}{2}\ln|\bm{\Lambda}_\beta| + \sum_{j=1}^J - \frac{1}{2} \mathrm{tr}\left(\left[\bm{\mu}_{\alpha,j}\bm{\mu}_{\alpha,j}^T + \bm{\Lambda_{\alpha,j}}\right] \left[\bm{I} \otimes E_{q(\bm{\Sigma}_j)}\left[\bm{\Sigma}_j^{-1}\right]\right]\right) + \frac{1}{2} \ln|\bm{\Lambda}_{\alpha,j}| + \cdots
\end{split}
\end{equation}

Updating $q(\bm{\beta})$ and $q(\bm{\alpha}_j)$ is more challenging because of the dependence of the weights $\bm{w}$ on those variational distributions. \cite{wand2014fully} proposes non-conjugate variational message passing that can be adapted to this case; he presents an algorithm for a Poisson hierarchical model with a single random effect. This can be adapted into a cyclical algorithm for arbitrary $J$. Algorithm~\ref{alg:CAVI} reports how to update $q(\bm{\beta})$ and $q(\bm{\alpha})$.

\begin{algorithm}[!ht]
	\caption{CAVI Updates for $q(\bm{\beta})$ and $q(\bm{\alpha})$}
	\label{alg:CAVI}
	\begin{algorithmic}	
		\State{\textbf{Initialize Variational Parameters}: Assume that $\{q(\bm{\Sigma}_j)\}_{j=1}^J$ are known and define the initial values of the variational parameters as $\bm{\mu}^{(0)}_{\bm{\beta}}$, $\bm{\Lambda}^{(0)}_{\bm{\beta}}$ for $q(\bm{\beta})$ and $\{\bm{\mu}^{(0)}_{\alpha,j}, \bm{\Lambda}^{(0)}_{\alpha,j}\}_{j=1}^J$  for $\{q(\bm{\alpha}_j)\}_{j=1}^J$}
		
		\vspace{1em}
		
		\State{0. Compute $\bm{w}$. Define $\bm{W} = \mathrm{diag}(\bm{w})$.}
		
		$$\ln \bm{w}^{(0)} = \bm{X}\bm{\mu}^{(0)}_{\bm{\beta}} + \bm{Z}\bm{\mu}^{(0)}_{\bm{\alpha}} + \frac{1}{2} \mathrm{diagonal}\left(\bm{X}\bm{\Lambda}^{(0)}_{\bm{\beta}}\bm{X}^T\right) + \sum_{j=1}^J \frac{1}{2} \mathrm{diagonal}\left(\bm{Z}_j\bm{\Lambda}^{(0)}_{\alpha,j}\bm{Z}_j^T\right)$$
		
		\State{1. Update $q(\bm{\beta})$}
		
		$$\bm{\Lambda}_{\bm{\beta}}^{*} = \left(\bm{X}^T \bm{W}^{(0)} \bm{X}\right)^{-1}; \quad \bm{\mu}_{\bm{\beta}}^{*} = \bm{\mu}_{\bm{\beta}}^{(0)} + \bm{\Lambda}_{\bm{\beta}}^{*}\bm{X}^T\left[\bm{y} - \bm{w}^{(0)}\right] $$
		
		$$\ln \bm{w}^{(1)} = \ln \bm{w}^{(0)} + \bm{X}\left(\bm{\mu}_{\bm{\beta}}^* - \bm{\mu}_{\bm{\beta}}^{(0)}\right) + \frac{1}{2} \mathrm{diagonal}\left(\bm{X} \left[\bm{\Lambda}^*_{\bm{\beta}} - \bm{\Lambda}^{(0)}_{\bm{\beta}}\right]\bm{X}^T\right)$$
		\For{$j$ in $\{1, \cdots, J\}$}
		
		\vspace{0.5em}
		
		\State{2.j: Update $q(\bm{\alpha}_j)$. Define $\bm{S}_j = \left(\bm{I} \otimes E_{q(\bm{\Sigma}_j)}[\bm{\Sigma}_j^{-1}]\right)$}
		
		$$\bm{\Lambda}_{\alpha, j}^{*} = \left(\bm{Z}_j^T \bm{W}^{(j)} \bm{Z}_j + \bm{S}_j\right)^{-1}; \quad \bm{\mu}_{\alpha,j}^{*} = \bm{\mu}_{\alpha,j}^{(0)} + \bm{\Lambda}_{\alpha,j}^{*}\left(\bm{Z}_j^T\left[\bm{y} - \bm{w}^{(j)}\right] - \bm{S}_j \bm{\mu}^{(0)}_{\alpha,j}\right) $$
		$$\ln \bm{w}^{(j+1)} = \ln \bm{w}^{(j)} + \bm{Z}_j\left(\bm{\mu}^*_{\alpha,j} - \bm{\mu}_{\alpha,j}^{(0)}\right) + \frac{1}{2} \mathrm{diagonal}\left(\bm{X} \left[\bm{\Lambda}^*_{\alpha,j} - \bm{\Lambda}^{(0)}_{\alpha,j}\right]\bm{X}^T\right)$$
		\EndFor
		
	\end{algorithmic}
\end{algorithm}

This algorithm is not guaranteed to increase the objective at each iteration; to address this, we fall back to ``damped'' updates (\citealt{knowles2011non}) if necessary. The updates for $q(\bm{\beta})$ and $q(\bm{\alpha})$, the updates for $q(\{\bm{\Sigma}_j\}_{j=1}^J)$ follow existing research (e.g., \citealt{wand2014fully}). As proposed in \citet{goplerud2023reeval}, all priors on $\bm{\Sigma}_j$ are weakly informative. In terms of including fixed effects, one can use an update for $q(\bm{\alpha}_j)$ where $\bm{S}_j = \bm{0}$ and the constraints are addressed as noted in Appendix~\ref{app:VI_FE}.

\section{Other Methodologies for Predicting Categorical Variables}
\label{app:other_methods}

This section provides additional details on other methods for predicting categorical outcomes. Specifically, it first discusses the ``One-vs-All'' approach (see, e.g., \citealt{rifkin2004defense}) for a review. Second, it notes some additional details about two stage models.

\subsection{Separate Regressions and One-vs-All}\label{app:separate_reg}

The main manuscript discussed an approach for estimating categorical outcomes using separate regressions following \citet{taddy2015distributed}'s work. This approach plugs in an estimate of the fixed effect, $\hat{\gamma}_i$, and then estimates separate models. In our setting, $\hat{\gamma}_i = 0$ and thus it is equivalent to ignoring the fixed effect entirely. After fitting $\bar{L}$ separate Poisson regressions, one obtains an estimate of the linear predictor $\hat{\psi}_{i,\ell} = \hat{\gamma}_i + \bm{z}_{i,\ell}^T \hat{\bm{\theta}}_\ell$ for each response category $\ell$ where we use $\bm{z}_{i,\ell}$ and $\bm{\theta}_\ell$ as a placeholder for the design matrix and coefficients for response category $\ell$. To obtain a prediction, these are combined in the usual ``softmax'' formulation for a multinomial logistic regression, i.e. $\hat{p}_{i,\ell} = \exp(\hat{\psi}_{i,\ell})/\sum_{k=1}^{\bar{L}} \exp(\hat{\psi}_{i,k})$. This can be done in a numerically stable way even if all $\hat{\psi}_{i,\ell}$ are small---see Appendix~\ref{app:software_demo} for an illustration.

In the remainder of this subsection, we note a close but under-appreciated connection between \citet{taddy2015distributed}'s approach and the popular approach ``One-vs-All'' in machine learning for multi-class classification, i.e. predicting a categorical outcome (see \citealt{bishop2006ml} for a textbook treatment and \citealt{rifkin2004defense} for a more in-depth discussion of this and other methods). At its core, One-vs-All argues that instead of trying to estimate a complex model for all categories, one can estimate $\bar{L}$ separate \emph{binary} classifiers. For each response category $\ell \in \bar{L}$, one predicts whether observation $y_i=\ell$ (or not)---hence the name ``one versus all'' or ``one versus rest.'' This produces a collection of probabilities $\{p^{\texttt{OvA}}_{i,\ell}\}$ that are generally normalized to sum to one using a ratio, i.e. $p_{i,\ell} = p^{\texttt{OvA}}_{i,\ell}/\sum_k p^{\texttt{OvA}}_{i,k}$, although more sophisticated methods exist (\citealt{rifkin2004defense}). 

The connection to \cite{taddy2015distributed}'s approach is clear: His method fits a Poisson regression where the outcomes are either zero or one, i.e., $I(y_i = \ell)$; One-vs-All would use a logistic regression. In general, one would expect the two methods to give very similar predictions. Thus, One-vs-All can be seen as one additional approximation beyond \citet{taddy2015distributed}'s approach to allow the use of any binary classifier versus a Poisson regression. In our numerical simulations in Section~\ref{app:validation_separate}, we thus compare both binary and Poisson regressions to see if One-vs-All versus \cite{taddy2015distributed}'s approach yields better results, although we do not expect much difference.

Finally, Section~\ref{sec:separate_reg} notes that one could partially pool either \cite{taddy2015distributed}'s approach or One-vs-All. While this is perhaps infeasible if $\bar{L}$ is enormous---although one might partially pool across some subset of $\bar{L}$ response categories, we suspect it will be rather important for performance given the limitsed data available for MRP and the ability to include alternative-specific covariates. To formally outline how this would work, assume that one estimated separate logistic regressions with linear predictor $\psi_{i,\ell} = \bm{z}_{i,l}^T \bm{\theta}_\ell$. The partially pooled One-vs-Rest would put a hierarchical prior on these coefficients $\bm{\theta}_\ell$ and estimate a single model---albeit without the fixed effect that is required to make it \emph{exactly} equivalent to multinomial regression. To run this in code, one simply takes the the model in the main text for \texttt{mvMRP} (see also Appendix~\ref{app:software_demo}) and simply removes the fixed effect, i.e. \verb@v_fe(case_id)@.

\color{black}
We note that this partially pooled separate regressions approach is somewhat similar in spirit to recent work by \citet{marble2025improving}. In their paper, a key methodological innovation to obtain the \emph{marginal} distribution of opinion on $J$ questions by fitting a partially pooled model---i.e. stack marginal responses to $J$ questions together and fit a model with random slopes on, e.g., ``age'' by question. The fundamental distinction from our approach is that they do not estimate the joint opinion on multiple questions. Rather, they seek to gain information from partially pooling across marginal models and use with auxiliary information on known results to calibrate MRP. We suspect their calibration could be applied to \texttt{mvMRP} and this is an interesting question for future research.
\color{black}

\subsection{Two-Stage MRP}\label{app:twostage_equiv}

This section briefly illustrates the equivalence of the two-stage procedure and the single multinomial logistic regression in a simple setting. In the example in Section~\ref{sec:twostep_validation}, the two-step model uses the following generative models for (i) a multinomial logistic regression to predict party using a demographic such as race and (ii) a logistic regression to predict policy choice using race and party. They are shown below.

$$\mathrm{Pr}(\texttt{partyID}_i = \ell~) \propto \exp(\beta^{\texttt{Stage1}}_{0,\ell} + \alpha^{\texttt{Stage1}}_{g[i],\ell})$$
$$\mathrm{Pr}(\texttt{policy}_i = 1~|~\texttt{partyID}_i, \bm{z}_i) \propto \exp(\alpha^{\texttt{Stage2}}_{g[i]} + \xi_{\texttt{partyID}_i})$$

Consider the multinomial logistic formulation:

$$\mathrm{Pr}(\texttt{partyID}_i = \ell, \mathrm{policy}_i = \ell' | \bm{z}_i) \propto \exp\left(\beta_{0,(\ell,\ell')} + \alpha_{g[i],(\ell,\ell')}\right)$$

In the unregularized model, the multinomial logistic regression generates predictions that correspond to the share of each combination of party and policy $(\ell,\ell')$ found in the estimation data as it is a fully saturated model. In the original two-stage model, this does not occur as the second stage model, i.e. predicting policy conditional on party, assumes that the effect of party is the same across racial groups. Thus, \emph{interacting} party and race together in the second model is necessary to create a fully saturated model. If one thus replaced the second model in the two-stage MRP with the following, the predictions of the two-stage and single multinomial logistic regression are identical.

$$\mathrm{Pr}(\texttt{policy}_i = 1~|~\texttt{partyID}_i,~\bm{z}_i) \propto \exp\left(\alpha^{\texttt{Stage2}^*}_{g[i], \texttt{partyID}_i} + \xi_{\texttt{partyID}_i}\right)$$

In the case where random effects or multiple demographic predictors are used, this exact equivalence does not hold. However, there are two implications for applied research: If one is performing Two-Stage MRP, one should interact party and demographics in the second stage model as this captures dependencies that a multinomial logistic regression would capture but an additive two stage model would miss. Second, standard uses of Two-Stage MRP can be seen as an approximation to multivariate MRP where a simpler model is assumed for the second stage. This both gives a solid footing to \texttt{mvMRP} as estimating the ``true'' quantity of interest (the joint distribution of party and policy) as well as giving a more formal justification to Two-Stage MRP.

\section{Validation}\label{app:validation}

This section contains additional information on the validation of \texttt{mvMRP} in Section~\ref{sec:twostep_validation} in the main text. It first describes the data from the Cooperative Election Study (CES) and post-stratification weights in more details and then proceeds to compare \texttt{mvMRP} against other specifications or ways of presenting the results.

\subsection{Additional Details on CES}\label{app:validation_CCES}

This section first outlines the exact question wording, with differences noted, for each policy used in our analysis. Survey data were downloaded from the CES (\citealt{dagonel2023cces}).

\begin{itemize}
	\item Abortion---20 Weeks (2016-20): ``Do you support or oppose each of the following proposals? [Ban ('18); Prohibit ('16,'20)] abortions after the 20th week of pregnancy.''
	\item Abortion---Ban (2016-20): ``Do you support or oppose each of the following proposals? Make abortion illegal in all circumstances.''
	\item Ban Assault Weapons (2016-20): ``On the issue of gun regulation, do you support or oppose each of the following proposals? Ban assault rifles.''
	\item Same-sex Marriage (2012-16): ``Do you favor or oppose allowing gays and lesbians to marry legally?''
	\item Immigration (2012-16): ``What do you think the U.S. government should do about immigration? Select all that apply... Grant legal status to all illegal immigrants who have held jobs and paid taxes for at least 3 years, and not been convicted of any felony crimes.''
	\item Renewable Energy (2016-20): ``Do you support or oppose each of the following proposals? Require that each state use a minimum amount of renewable fuels (wind, solar, and hydroelectric) in the generation of electricity even if electricity prices increase [somewhat ('16); a little ('20)].''
\end{itemize}

For post-stratification, we aggregate 5-year American Community Survey microdata from IPUMS USA (\citealt{ruggles2025ipums}) to construct the joint distribution of the following variables: sex (2 levels); age (6 levels: 18-29, 30-39, 40-49, 50-59, 60-69, 70+); race (5 levels: white, Black, Latino, Asian, other); education (4 levels: no high school diploma, high school, some college, college degree). For questions occurring in the CES from 2012-16, microdata are from the 2016 ACS; for those occurring 2016-20, microdata are from the 2019 ACS.\footnote{Ideally, we would have used the 2020 ACS, sampled over the period 2016-2020. However, the Covid-19 pandemic disrupted the Census Bureau's ACS data collection in 2020; see \url{https://www.census.gov/newsroom/blogs/random-samplings/2021/10/pandemic-impact-on-2020-acs-1-year-data.html}.} The poststratification frame also includes state-level Democratic presidential vote share in 2012/2016 and state percent Evangelical measured in 2010 from the Association of Religion Data Archives.

In terms of sample size, the main text notes that the median state/party combination has around 750 respondents. This is computing using the survey weights, as is all of our ground truth estimates from the superpoll. The lower 25th percentile is around 359 respondents. In terms of respondents per state, the median state has 2480 respondents, with the smallest state having 273 respondents. Thus, our superpool is large enough for us to reliably estimate the joint distribution of the questions as well as downstream quantites such as support for a policy by party.

\subsection{Raw Mean Absolute Error for Main Models}\label{app:validation_raw_mae}

Table~\ref{tab:app_raw_error} reports the median of the raw mean absolute error across 500 simulations (rather than the percentage change shown in the main text). This figure shows the enormous error of the ``Naive'' method for estimating conditional distributions (e.g., around 13 percentage points on average). For all other questions and quantities, Two-Stage (``2S'') and both variants of \texttt{mvMRP} (\texttt{mvMRP} and Copart.) are within normal error rates seen for MRP at the state level, i.e. a MAE between 2 and 5 percentage points.

\begin{table}[!htbp]
	\caption{Raw Mean Absolute Error}
	\label{tab:app_raw_error}
	\begin{centering}
		\begin{tabular}{r@{\,}r|r|r|r|r|r|r|r}
			& \multicolumn{1}{c|}{Policy} &  Naive & Taddy & 2S & \texttt{mvMRP} & Copart. \\
			\hline\hline
			\ldelim\{{7}{*}[Joint~] & Abortion (Ban) & 0.0280 & 0.0340 & 0.0270 & 0.0270 & 0.0240 \\
 & Abortion (20 Weeks) & 0.0460 & 0.0350 & 0.0280 & 0.0270 & 0.0240 \\
 & Assault Weapons & 0.0520 & 0.0360 & 0.0280 & 0.0280 & 0.0260 \\
 & Gay Marriage & 0.0500 & 0.0360 & 0.0290 & 0.0300 & 0.0270 \\
 & Immigration & 0.0430 & 0.0330 & 0.0270 & 0.0270 & 0.0240 \\
 & Renewable Energy & 0.0480 & 0.0350 & 0.0280 & 0.0290 & 0.0260 \\
 & \textbf{Average} & \textbf{0.0445} & \textbf{0.0348} & \textbf{0.0278} & \textbf{0.0280} & \textbf{0.0252} \\
\ldelim\{{7}{*}[Conditional~] & Abortion (Ban) & 0.0750 & 0.0450 & 0.0360 & 0.0360 & 0.0360 \\
 & Abortion (20 Weeks) & 0.1280 & 0.0470 & 0.0390 & 0.0370 & 0.0370 \\
 & Assault Weapons & 0.1660 & 0.0540 & 0.0410 & 0.0410 & 0.0420 \\
 & Gay Marriage & 0.1560 & 0.0650 & 0.0470 & 0.0460 & 0.0460 \\
 & Immigration & 0.1260 & 0.0400 & 0.0400 & 0.0380 & 0.0380 \\
 & Renewable Energy & 0.1480 & 0.0490 & 0.0440 & 0.0430 & 0.0430 \\
 & \textbf{Average} & \textbf{0.1332} & \textbf{0.0500} & \textbf{0.0412} & \textbf{0.0402} & \textbf{0.0403} \\
\ldelim\{{7}{*}[\shortstack{Party\\Marginal}~] & Abortion (Ban) & 0.0400 & 0.0590 & 0.0460 & 0.0460 & 0.0400 \\
 & Abortion (20 Weeks) & 0.0400 & 0.0600 & 0.0470 & 0.0470 & 0.0390 \\
 & Assault Weapons & 0.0410 & 0.0580 & 0.0460 & 0.0460 & 0.0390 \\
 & Gay Marriage & 0.0400 & 0.0540 & 0.0470 & 0.0490 & 0.0400 \\
 & Immigration & 0.0400 & 0.0590 & 0.0470 & 0.0480 & 0.0400 \\
 & Renewable Energy & 0.0410 & 0.0590 & 0.0470 & 0.0470 & 0.0400 \\
 & \textbf{Average} & \textbf{0.0403} & \textbf{0.0582} & \textbf{0.0467} & \textbf{0.0472} & \textbf{0.0397} \\
\ldelim\{{7}{*}[\shortstack{Policy\\Marginal}~] & Abortion (Ban) & 0.0290 & 0.0420 & 0.0290 & 0.0310 & 0.0310 \\
 & Abortion (20 Weeks) & 0.0330 & 0.0590 & 0.0340 & 0.0340 & 0.0350 \\
 & Assault Weapons & 0.0380 & 0.0730 & 0.0400 & 0.0430 & 0.0410 \\
 & Gay Marriage & 0.0390 & 0.0670 & 0.0390 & 0.0410 & 0.0430 \\
 & Immigration & 0.0300 & 0.0380 & 0.0300 & 0.0280 & 0.0280 \\
 & Renewable Energy & 0.0380 & 0.0550 & 0.0400 & 0.0400 & 0.0390 \\
 & \textbf{Average} & \textbf{0.0345} & \textbf{0.0557} & \textbf{0.0353} & \textbf{0.0362} & \textbf{0.0362}
\\\hline\hline
		\end{tabular}
		\caption*{\footnotesize \emph{Note:} This table shows the mean absolute error across states as discussed in the main text; the number is the median across 500 simulations.}
	\end{centering}
\end{table}

\subsection{(Partially) Separate Regressions}\label{app:validation_separate}

This section compares the performance of many different methods for either separate regressions or partially pooled One-vs-All (see Appendix~\ref{app:separate_reg}). We show performance versus Two-Stage MRP for comparability to the figures in the main text.

We explore wholly separate regressions (``One-vs-All'') with regularization based on random effects, i.e. fitting $\bar{L}$ separate hierarchical models. We examine this for both binomial regressions (``Bin.'')---a more traditional ``One-vs-All'' as well as a Poisson regression (``Pois.''), i.e., adapting \citealt{taddy2015distributed}'s framework but using random effects instead of his preferred method of gamma lasso regression. We also examine the partially pooled One-vs-All models (PP-OvA); Poisson PP-OvA is identical to \texttt{mvMRP} with the fixed effect for individual removed. Binomial PP-OvA uses a binomial likelihood instead of a Poisson likelihood for the same model. Predictions are produced as discussed in Appendix~\ref{app:separate_reg}. Finally, we again report the results from Taddy's \texttt{distrom} package using gamma lasso regressions as well as a manual implementation that adds lagged copartisanship as \texttt{distrom} does not allow alternative specific covariates.

Table~\ref{tab:app_sep} shows the impact on MAE for joint, conditional and both marginal distributions as in the main text where the reference category is again Two-Stage MRP to ensure comparability with the results in the main text. Thus, the column ``Copart.'' exactly replicates Table~\ref{tab:validate_joint}  and Table~\ref{tab:validate_marginal} in the main text.

\begin{table}[!htbp]
	\caption{Validation on Separate Regressions}
	\label{tab:app_sep}
	\begin{centering}
		\resizebox{\textwidth}{!}{%
			\begin{tabular}{r@{\,}r|r|r|r|r|r|r|r}
				\hline\hline
				& \multicolumn{1}{c|}{Policy} & \multicolumn{1}{c}{Copart.} & \multicolumn{2}{c}{One-vs-All} & \multicolumn{2}{c}{PP-OvA} & \multicolumn{2}{c}{Taddy} \\
				& & & Bin. & Pois. & Bin. &Pois. & T & CT \\
				\ldelim\{{7}{*}[Joint~] & Abortion (Ban) & -8.04 & 12.15 & 15.44 & -8.89 & -7.70 & 26.93 & 28.51 \\
 & Abortion (20 Weeks) & -12.42 & 12.14 & 13.39 & -13.11 & -11.91 & 28.50 & 29.37 \\
 & Assault Weapons & -7.32 & 10.91 & 14.52 & -7.02 & -5.90 & 28.79 & 30.81 \\
 & Gay Marriage & -7.93 & 9.20 & 12.13 & -7.34 & -7.35 & 24.43 & 25.55 \\
 & Immigration & -12.24 & 13.52 & 16.61 & -12.20 & -11.80 & 21.41 & 21.42 \\
 & Renewable Energy & -9.60 & 9.63 & 12.75 & -10.53 & -9.42 & 23.19 & 24.38 \\
 & \textbf{Average} & \textbf{-9.59} & \textbf{11.26} & \textbf{14.14} & \textbf{-9.85} & \textbf{-9.01} & \textbf{25.54} & \textbf{26.67} \\
\ldelim\{{7}{*}[Cond.~] & Abortion (Ban) & 0.43 & 13.89 & 19.18 & -0.27 & 0.01 & 24.16 & 24.10 \\
 & Abortion (20 Weeks) & -4.47 & 19.95 & 21.38 & -3.81 & -4.61 & 20.19 & 21.21 \\
 & Assault Weapons & 2.04 & 15.29 & 19.50 & 3.29 & 3.15 & 32.46 & 32.71 \\
 & Gay Marriage & -1.96 & 12.91 & 15.21 & -0.30 & -1.16 & 38.48 & 40.75 \\
 & Immigration & -4.93 & 24.81 & 27.29 & -5.05 & -5.17 & -0.50 & -0.43 \\
 & Renewable Energy & -3.44 & 11.38 & 13.53 & -5.17 & -3.52 & 10.94 & 10.41 \\
 & \textbf{Average} & \textbf{-2.05} & \textbf{16.37} & \textbf{19.35} & \textbf{-1.89} & \textbf{-1.88} & \textbf{20.95} & \textbf{21.46} \\
\ldelim\{{7}{*}[Party~] & Abortion (Ban) & -13.66 & 9.55 & 11.71 & -14.14 & -12.81 & 27.41 & 29.38 \\
 & Abortion (20 Weeks) & -16.14 & 9.60 & 10.95 & -16.55 & -15.40 & 27.70 & 28.67 \\
 & Assault Weapons & -15.32 & 9.91 & 12.41 & -15.18 & -14.16 & 24.97 & 26.81 \\
 & Gay Marriage & -14.76 & 8.85 & 11.48 & -14.95 & -13.99 & 15.17 & 15.40 \\
 & Immigration & -15.61 & 8.99 & 11.08 & -15.55 & -15.05 & 25.96 & 25.49 \\
 & Renewable Energy & -14.52 & 10.00 & 12.70 & -15.05 & -13.64 & 26.21 & 28.04 \\
 & \textbf{Average} & \textbf{-15.00} & \textbf{9.48} & \textbf{11.72} & \textbf{-15.24} & \textbf{-14.18} & \textbf{24.57} & \textbf{25.63} \\
\ldelim\{{7}{*}[Policy~] & Abortion (Ban) & 7.99 & 5.61 & 16.57 & 9.29 & 8.38 & 44.18 & 44.44 \\
 & Abortion (20 Weeks) & 1.07 & 3.00 & 1.24 & -0.14 & 2.52 & 71.24 & 73.46 \\
 & Assault Weapons & 1.86 & 7.12 & 15.20 & 5.08 & 7.01 & 82.86 & 84.67 \\
 & Gay Marriage & 8.79 & -0.83 & 2.55 & 9.38 & 10.64 & 69.87 & 78.86 \\
 & Immigration & -5.38 & 11.18 & 12.89 & -4.87 & -3.93 & 25.45 & 25.11 \\
 & Renewable Energy & -2.45 & 1.79 & 8.77 & -1.80 & -0.78 & 39.45 & 41.57 \\
 & \textbf{Average} & \textbf{1.98} & \textbf{4.65} & \textbf{9.54} & \textbf{2.82} & \textbf{3.97} & \textbf{55.51} & \textbf{58.02}
\\\hline\hline
			\end{tabular}
		}
		\caption*{\footnotesize \emph{Note:} This table shows the percentage change in mean absolute error across states as discussed in the main text. Negative numbers indicate the method out-performs Two-Stage MRP. ``T.'' stands for ``Taddy'' without copartisanship, i.e. the model in the main text, and ``CT'' stands for ``Copart. + Taddy'', i.e. adding copartisanship. ``Cond.'' stands for ``conditional'', ``Party'' refers to the marginal distribution of opinion on partisanship and ``Policy'' refers to the marginal distribution of opinion on policy.}
	\end{centering}
\end{table}

Focusing on the joint and conditional distributions, it shows that pooling across issues, fully separate regressions (either via One-vs-All or Taddy) perform unacceptably poorly. Besides doing much worse than standard Two-Stage MRP, they do about 20\% worse than \texttt{mvMRP} with copartisanship. By contrast, if we use partial pooling in conjunction with the One-vs-All approach, we get comparable performance to \texttt{mvMRP}. This suggests that while including the fixed effect allows us to represent the problem as a well-grounded statistical model, excluding this variable does not materially hurt performance. A similar story appears when we examine the marginal distributions, although the partially pooled One-vs-All models do slightly worse on the estimation of the marginal distribution of policy opinion.

\subsection{Pooling Across Simulations}\label{app:validation_mean_vs_median}

In the main text, we consider the median across 500 simulations of the mean absolute error for each method. Table~\ref{tab:app_mean_ratio} show the average of the mean absolute error across simulations and report the percentage change versus Two-Stage MRP. \texttt{mvMRP} looks somewhat stronger in this approach, especially for policy marginal where it now performs about equally to Two-Stage MRP (vs. slightly worse). This is because there are (rare) instances where \texttt{mgcv} insufficiently regularizes the estimates for Two-Stage MRP and results in quite poor estimates. 

\begin{table}[!htbp]
	\caption{Validation Using Average Pooling Across Simulations}
	\label{tab:app_mean_ratio}
	\begin{centering}
		\begin{tabular}{r@{\,}r|r|r|r|r}
			& \multicolumn{1}{c|}{Policy} &  Naive & Taddy &  \texttt{mvMRP} & Copart. \\
			\hline\hline
			\ldelim\{{7}{*}[Joint~] & Abortion (Ban) & 6.45 & 26.07 & 0.46 & -7.99 \\
 & Abortion (20 Weeks) & 67.34 & 27.98 & -0.45 & -11.89 \\
 & Assault Weapons & 84.96 & 26.54 & -0.54 & -9.36 \\
 & Gay Marriage & 65.32 & 20.58 & -0.61 & -10.13 \\
 & Immigration & 55.38 & 20.62 & -1.08 & -12.29 \\
 & Renewable Energy & 66.76 & 22.93 & 0.07 & -9.41 \\
 & \textbf{Average} & \textbf{57.70} & \textbf{24.12} & \textbf{-0.36} & \textbf{-10.18} \\
\ldelim\{{7}{*}[Conditional~] & Abortion (Ban) & 103.94 & 23.07 & -0.10 & -0.30 \\
 & Abortion (20 Weeks) & 221.06 & 19.49 & -4.97 & -4.78 \\
 & Assault Weapons & 283.59 & 26.24 & -2.72 & -1.00 \\
 & Gay Marriage & 208.66 & 29.27 & -7.45 & -7.41 \\
 & Immigration & 210.03 & -1.49 & -5.07 & -4.97 \\
 & Renewable Energy & 227.64 & 10.01 & -3.85 & -3.59 \\
 & \textbf{Average} & \textbf{209.15} & \textbf{17.76} & \textbf{-4.03} & \textbf{-3.67} \\
\ldelim\{{7}{*}[\shortstack{Party\\Marginal}~] & Abortion (Ban) & -12.88 & 26.72 & 0.09 & -13.46 \\
 & Abortion (20 Weeks) & -13.04 & 28.18 & 0.77 & -15.06 \\
 & Assault Weapons & -12.65 & 23.82 & -0.49 & -15.47 \\
 & Gay Marriage & -14.43 & 14.30 & 3.37 & -14.44 \\
 & Immigration & -14.89 & 25.63 & 0.71 & -15.56 \\
 & Renewable Energy & -12.92 & 26.01 & 0.59 & -14.17 \\
 & \textbf{Average} & \textbf{-13.47} & \textbf{24.11} & \textbf{0.84} & \textbf{-14.69} \\
\ldelim\{{7}{*}[\shortstack{Policy\\Marginal}~] & Abortion (Ban) & -1.93 & 41.68 & 6.52 & 6.87 \\
 & Abortion (20 Weeks) & -4.05 & 65.47 & -2.51 & -0.39 \\
 & Assault Weapons & -6.84 & 71.74 & 2.91 & -2.29 \\
 & Gay Marriage & -6.30 & 59.20 & -1.00 & 2.58 \\
 & Immigration & -1.39 & 19.27 & -6.56 & -5.76 \\
 & Renewable Energy & -3.96 & 36.77 & 1.14 & -2.77 \\
 & \textbf{Average} & \textbf{-4.08} & \textbf{49.02} & \textbf{0.08} & \textbf{-0.29}
\\\hline\hline
		\end{tabular}
		\caption*{\footnotesize \emph{Note:} This table shows the percentage change in mean absolute error across states as discussed in the main text where the variability across simulations is addressed by using the average MAE. Negative numbers indicate the method out-performs Two-Stage MRP.}
	\end{centering}
\end{table}

\subsection{Other Specifications with Copartisanship}\label{app:validation_copart}

We next consider different specifications that include copartisanship. We consider first adding copartisanship to the \texttt{mgcv} estimation of Two-Stage MRP (``2S + Copart.''). Second, we consider estimating a two-stage model using variational inference to check that this approach works well regardless of the inferential procedure chosen (``Two-Stage VI''). We also include \texttt{mvMRP} with copartianship as a reference. Table~\ref{tab:app_copart} shows the relative change in MAE with Two-Stage as the reference category. Thus, the column ``Copart. '' exactly replicates the main results in Tables~\ref{tab:validate_joint} and~\ref{tab:validate_marginal}. As expected, including copartianship in Two-Stage MRP improves performance on the joint distribution (by around 3\%) and the party marginal (by around 7\%). Interestingly, it results in slightly worse peformance for the conditional distribution. This could be because \texttt{mgcv} is limited in how alternative specific covariates can be included.

The final column considers Two-Stage MRP estimated using variational inference in each stage. The first stage, i.e. predicting partisanship, uses the same specification as \texttt{mvMRP} with lagged copartianship but now merely tries to post-stratify a single question---i.e. MRP with a single multinomial question as its outcome. We see this performs comparably to estimation with \texttt{mgcv} albeit with larger improvements for estimating the joint distribution. It performs slightly worse on the conditional but slightly better on the policy marginal. Thus, if Two-Stage MRP is required, using the methods in this paper to do VI for each stage would likely perform well.

\begin{table}[!htbp]
	\caption{Validation on Different Models with Copartianship}
	\label{tab:app_copart}
	\begin{centering}
		\begin{tabular}{r@{\,}r|r|r|r}
			\hline\hline
			& \multicolumn{1}{c|}{Policy} & \multicolumn{1}{c}{Copart.} & 2S + Copart. & Two-Stage VI \\
			\ldelim\{{7}{*}[Joint~] & Abortion (Ban) & -8.04 & -2.36 & -10.28 \\
 & Abortion (20 Weeks) & -12.42 & -3.23 & -11.42 \\
 & Assault Weapons & -7.32 & -2.72 & -8.58 \\
 & Gay Marriage & -7.93 & -4.02 & -9.71 \\
 & Immigration & -12.24 & -3.63 & -12.05 \\
 & Renewable Energy & -9.60 & -3.09 & -9.66 \\
 & \textbf{Average} & \textbf{-9.59} & \textbf{-3.17} & \textbf{-10.28} \\
\ldelim\{{7}{*}[Conditional~] & Abortion (Ban) & 0.43 & 1.80 & 0.86 \\
 & Abortion (20 Weeks) & -4.47 & 1.60 & 0.74 \\
 & Assault Weapons & 2.04 & 3.21 & 2.16 \\
 & Gay Marriage & -1.96 & 1.49 & 0.90 \\
 & Immigration & -4.93 & 2.92 & 2.59 \\
 & Renewable Energy & -3.44 & 2.95 & 2.65 \\
 & \textbf{Average} & \textbf{-2.05} & \textbf{2.33} & \textbf{1.65} \\
\ldelim\{{7}{*}[\shortstack{Party\\Marginal}~] & Abortion (Ban) & -13.66 & -5.63 & -16.19 \\
 & Abortion (20 Weeks) & -16.14 & -6.28 & -17.27 \\
 & Assault Weapons & -15.32 & -5.37 & -15.93 \\
 & Gay Marriage & -14.76 & -8.70 & -18.02 \\
 & Immigration & -15.61 & -7.78 & -18.65 \\
 & Renewable Energy & -14.52 & -6.15 & -16.59 \\
 & \textbf{Average} & \textbf{-15.00} & \textbf{-6.65} & \textbf{-17.11} \\
\ldelim\{{7}{*}[\shortstack{Policy\\Marginal}~] & Abortion (Ban) & 7.99 & -0.36 & 0.95 \\
 & Abortion (20 Weeks) & 1.07 & 1.29 & 0.40 \\
 & Assault Weapons & 1.86 & -2.98 & -9.27 \\
 & Gay Marriage & 8.79 & 3.05 & 1.09 \\
 & Immigration & -5.38 & 3.19 & 0.05 \\
 & Renewable Energy & -2.45 & -1.29 & -7.51 \\
 & \textbf{Average} & \textbf{1.98} & \textbf{0.48} & \textbf{-2.38}
\\\hline\hline
		\end{tabular}
		\caption*{\footnotesize \emph{Note:} This table shows the percentage change in mean absolute error across states as discussed in the main text. Negative numbers indicate the method out-performs Two-Stage MRP without copartisanship included.}
	\end{centering}
\end{table}

\subsection{Other Specifications of \texttt{mvMRP}}\label{app:validation_shallow}

Next, we consider other specifications of \texttt{mvMRP}. We look at a simpler specification (``Shallow'') that only includes interactions between the marginal questions and demographics---e.g. \verb@(1|age:policy)+(1|age:partyID)@---and not the combined question, i.e. 

\noindent\verb@(1|age:choice)@ or equivalently \verb@(1|age:policy:partyID)@. This is a simpler model so could perform well with more limited data. We also look at a more complex specification (``Deep'') that adds all two and three-way interactions between demographics (interacted with questions/response categories, e.g. \verb@(1|age:sex:policy)@) as well as interactions between geography (state and division) and demographics (again interacted with questions/response categories, e.g., \verb@(1|state:age:policy)@). This model takes longer to estimate---around one hour. We believe this is probably too complex of a model to estimate with only 2,000 respondents but, if one had more respondents, it could produce better performance. This model is thus shown to ensure that a ``deep'' does not have catastrophic performance at small sample sizes. \cite{goplerud2022mavb} suggests there should be some overfitting if too complex a model is used, and thus we suspect this may have slightly worse performance. For reference, we also include \texttt{mvMRP} with copartisanship. This column thus exactly replicates the tables in the main manuscript. Negative numbers indicates the method out-performs Two-Stage MRP. The results here are broadly as expected; the ``shallow'' model does slightly worse--especially in terms of estimating the conditional distributions, although it is still very comparable to Two-Stage MRP. The deep model also does slightly worse than our preferred specification, but is broadly comparable except---perhaps---on the estimation of the marginal opinion on policy where it is consistently weaker than both the ``Shallow'' and standard \texttt{mvMRP}.

\begin{table}[!htbp]
	\caption{Validation on Different \texttt{mvMRP} Specifications}
	\label{tab:app_mvmrp_spec}
	\begin{centering}
		\begin{tabular}{r@{\,}r|r|r|r}
			& \multicolumn{1}{c|}{Policy} & Copart. & Shallow & Deep \\
			\hline\hline
			\ldelim\{{7}{*}[Joint~] & Abortion (Ban) & -8.04 & -8.46 & -7.57 \\
 & Abortion (20 Weeks) & -12.42 & -11.50 & -11.17 \\
 & Assault Weapons & -7.32 & -7.11 & -7.33 \\
 & Gay Marriage & -7.93 & -6.47 & -7.48 \\
 & Immigration & -12.24 & -11.50 & -11.45 \\
 & Renewable Energy & -9.60 & -7.74 & -8.47 \\
 & \textbf{Average} & \textbf{-9.59} & \textbf{-8.80} & \textbf{-8.91} \\
\ldelim\{{7}{*}[Conditional~] & Abortion (Ban) & 0.43 & 0.71 & -0.16 \\
 & Abortion (20 Weeks) & -4.47 & -1.90 & -2.92 \\
 & Assault Weapons & 2.04 & 3.37 & 1.70 \\
 & Gay Marriage & -1.96 & 1.94 & -0.80 \\
 & Immigration & -4.93 & -4.14 & -3.97 \\
 & Renewable Energy & -3.44 & -0.25 & -3.37 \\
 & \textbf{Average} & \textbf{-2.05} & \textbf{-0.04} & \textbf{-1.59} \\
\ldelim\{{7}{*}[\shortstack{Party\\Marginal}~] & Abortion (Ban) & -13.66 & -13.98 & -12.56 \\
 & Abortion (20 Weeks) & -16.14 & -16.18 & -15.42 \\
 & Assault Weapons & -15.32 & -15.81 & -14.43 \\
 & Gay Marriage & -14.76 & -14.57 & -14.19 \\
 & Immigration & -15.61 & -15.20 & -14.76 \\
 & Renewable Energy & -14.52 & -14.48 & -13.51 \\
 & \textbf{Average} & \textbf{-15.00} & \textbf{-15.04} & \textbf{-14.14} \\
\ldelim\{{7}{*}[\shortstack{Policy\\Marginal}~] & Abortion (Ban) & 7.99 & 7.73 & 7.41 \\
 & Abortion (20 Weeks) & 1.07 & 1.67 & 6.03 \\
 & Assault Weapons & 1.86 & 2.68 & 4.21 \\
 & Gay Marriage & 8.79 & 8.09 & 11.15 \\
 & Immigration & -5.38 & -5.85 & -0.38 \\
 & Renewable Energy & -2.45 & -1.57 & 0.81 \\
 & \textbf{Average} & \textbf{1.98} & \textbf{2.12} & \textbf{4.87}
\\\hline\hline
		\end{tabular}
		\caption*{\footnotesize \emph{Note:} This table shows the percentage change in mean absolute error across states as discussed in the main text. Negative numbers indicate the method out-performs Two-Stage MRP.}
	\end{centering}
\end{table}

\section{Responsiveness Analyses}
\label{app:responsiveness}

This appendix provides additional information about the empirical application to multidimensional responsiveness in Section~\ref{sec:application} of the main text of the paper.

\subsection{Additional Information on Bills and Survey Questions}
\label{app:responsiveness_bills_surveys}

The multidimensional responsiveness analysis relies on roll-call votes from two bills considered by the U.S. Senate in the 113th Congress. We match these to questions on the 2014 CES, as follows:
\begin{itemize}
	\item Assault Weapons Ban of 2013 (S. 150): Failed on 40-60 vote. CES question: ``On the issue of gun regulation, are you for or against each of the following proposals? ... Ban assault rifles.''
	
	\item Agricultural Act of 2014 (H.R. 2642): Passed on 68-32 vote. CES question: ``Congress considered many important bills over the past two years. For each of the following tell us whether you support or oppose the legislation in principle ... Agriculture Bill: Ends price supports for corn, wheat, sugar and other agricultural products. Creates a federally subsidized crop insurance program. Reauthorizes the food stamp program, but cuts 10\% of the program's funding.'' 
\end{itemize}

\begin{table}[!htb]
	\caption{Senators' Roll-Call Votes}
	\small
	\label{tab:senvotes}
	\begin{centering}
		
\begin{tabular}{>{\centering\arraybackslash}m{0.3in}| >{\raggedright\arraybackslash}m{2.8in}>{\raggedright\arraybackslash}m{2.8in}}
\toprule
 \multicolumn{1}{c}{ }& \multicolumn{1}{c}{\textbf{Assault Ban: Yes Vote}} & \multicolumn{1}{c}{\textbf{Assault Ban: No Vote}}\\
  \midrule
  \rotatebox{90}{\textbf{Farm Bill: Yes Vote}} & \underline{Republicans:} Kirk (IL). \underline{Democrats:} Baldwin (WI), Boxer (CA), Brown (OH), Cantwell (WA), Cardin (MD), Carper (DE), Coons (DE), Durbin (IL), Feinstein (CA), Franken (MN), Harkin (IA), Hirono (HI), Kaine (VA), Klobuchar (MN), Leahy (VT), Levin (MI), McCaskill (MO), Menendez (NJ), Merkley (OR), Mikulski (MD), Murray (WA), Nelson (FL), Reid (NV), Rockefeller (WV), Schatz (HI), Schumer (NY), Shaheen (NH), Stabenow (MI), and Wyden (OR). \underline{Independents:} Sanders (VT). & \underline{Republicans:} Alexander (TN), Blunt (MO), Boozman (AR), Chambliss (GA), Coats (IN), Cochran (MS), Crapo (ID), Enzi (WY), Fischer (NE), Graham (SC), Hatch (UT), Hoeven (ND), Isakson (GA), Johanns (NE), McConnell (KY), Moran (KS), Portman (OH), Risch (ID), Thune (SD), Vitter (LA), and Wicker (MS). \underline{Democrats:} Baucus (MT), Begich (AK), Bennet (CO), Donnelly (IN), Hagan (NC), Heinrich (NM), Heitkamp (ND), Johnson (SD), Landrieu (LA), Manchin (WV), Pryor (AR), Tester (MT), Udall (CO), Udall (NM), and Warner (VA). \underline{Independents:} King (ME).\par\\
\rotatebox{90}{\textbf{Farm Bill: No Vote}} & \underline{Democrats:} Blumenthal (CT), Casey (PA), Gillibrand (NY), Murphy (CT), Reed (RI), Warren (MA), and Whitehouse (RI). & \underline{Republicans:} Ayotte (NH), Barrasso (WY), Burr (NC), Coburn (OK), Collins (ME), Corker (TN), Cornyn (TX), Cruz (TX), Flake (AZ), Grassley (IA), Heller (NV), Inhofe (OK), Johnson (WI), Lee (UT), McCain (AZ), Murkowski (AK), Paul (KY), Roberts (KS), Rubio (FL), Scott (SC), Sessions (AL), Shelby (AL), and Toomey (PA).\\
\bottomrule
\end{tabular}

		\caption*{\footnotesize \emph{Note:} This table shows how each senator voted across the two bills. We drop two senators who left the Senate between the votes on these bills, as well as their replacements---Mo Cowan and Ed Markey (MA), and Jeffrey Chiesa and Cory Booker (NJ).}
	\end{centering}
\end{table}

Table~\ref{tab:senvotes} reports senators' roll-call votes across the two bills. As noted in the main text, the modal Republican voted against both bills, while the modal Democrat voted for both. However, there is considerable variation both within and across parties. For example, 15 Democrats representing swing or traditionally Republican states voted for the Farm Bill and against the assault weapons ban, and seven (mostly progressives from the Northeast) voted for the assualt weapons ban but against the Farm Bill. On the Republican side, 21 senators voted for the Farm Bill but against the assault weapons ban. The most surprising vote of all, perhaps, was that of Senator Mark Kirk, Republican of Illinois, who represented a particularly liberal state.

Regarding post-stratification, we use the five-year 2014  ACS and code all variables as discussed in Appendix~\ref{app:validation_CCES} with the addition of household income as a five-category variable (under \$20k,
\$20k-\$40k, \$40k-\$80k, \$80k-\$150k, and \$150k+) following \cite{ghitza2013mrp}.

\subsection{Quantities of Interest}
\label{app:responsivness_qoi}

\begin{figure}[!htb]
	\centering
	\caption{Joint Support by State and Party}
	\vspace{-0.5em}
	\includegraphics[width=0.95\textwidth]{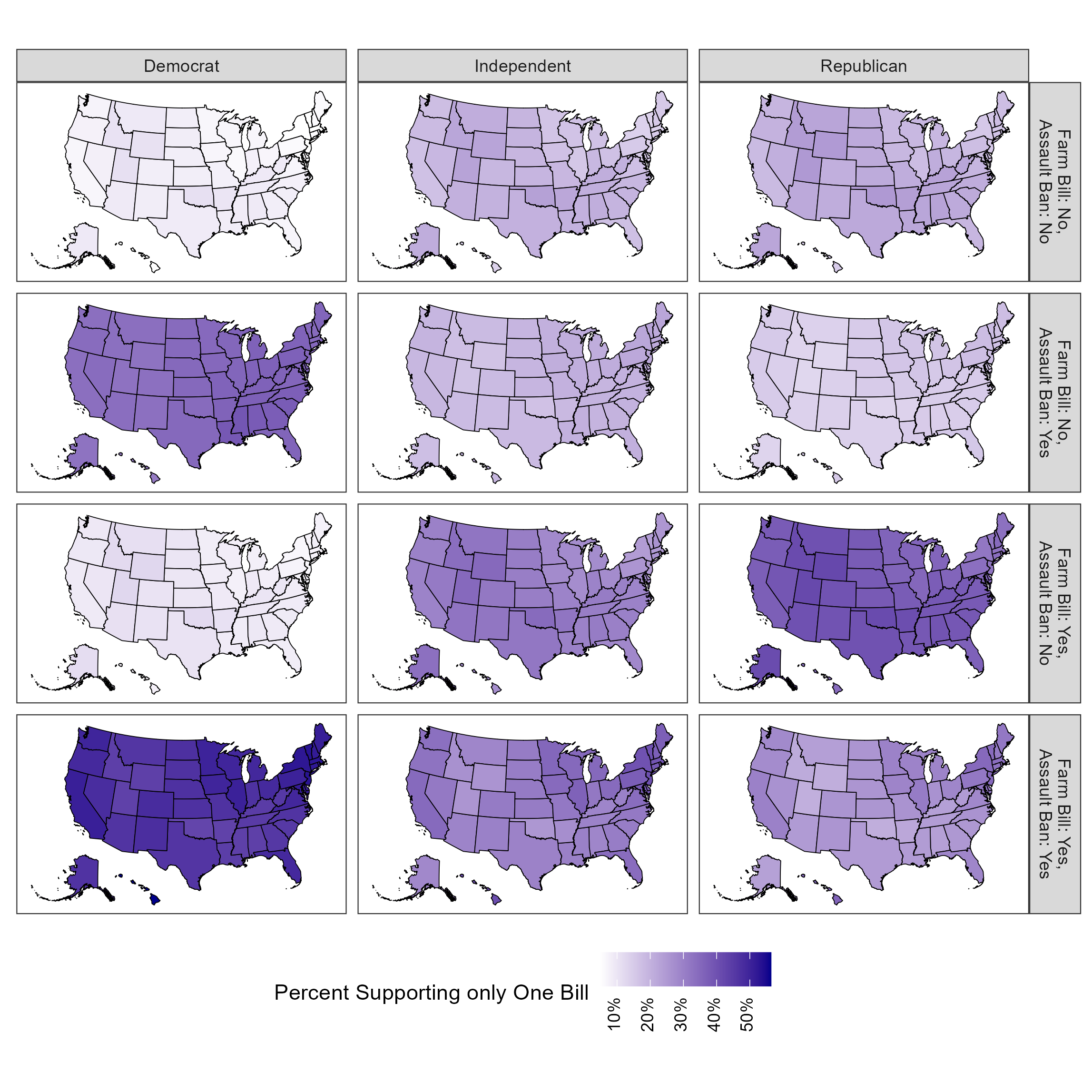}
	\label{fig:mapbyparty}
	\caption*{\footnotesize \emph{Note:} The figure shows the share of each state and party that support each combination of policies.}
\end{figure}

\begin{figure}[htb]
	\centering
	\caption{Heterogeneity of Opinion}
	\vspace{-0.5em}
	\includegraphics[width=0.95\textwidth]{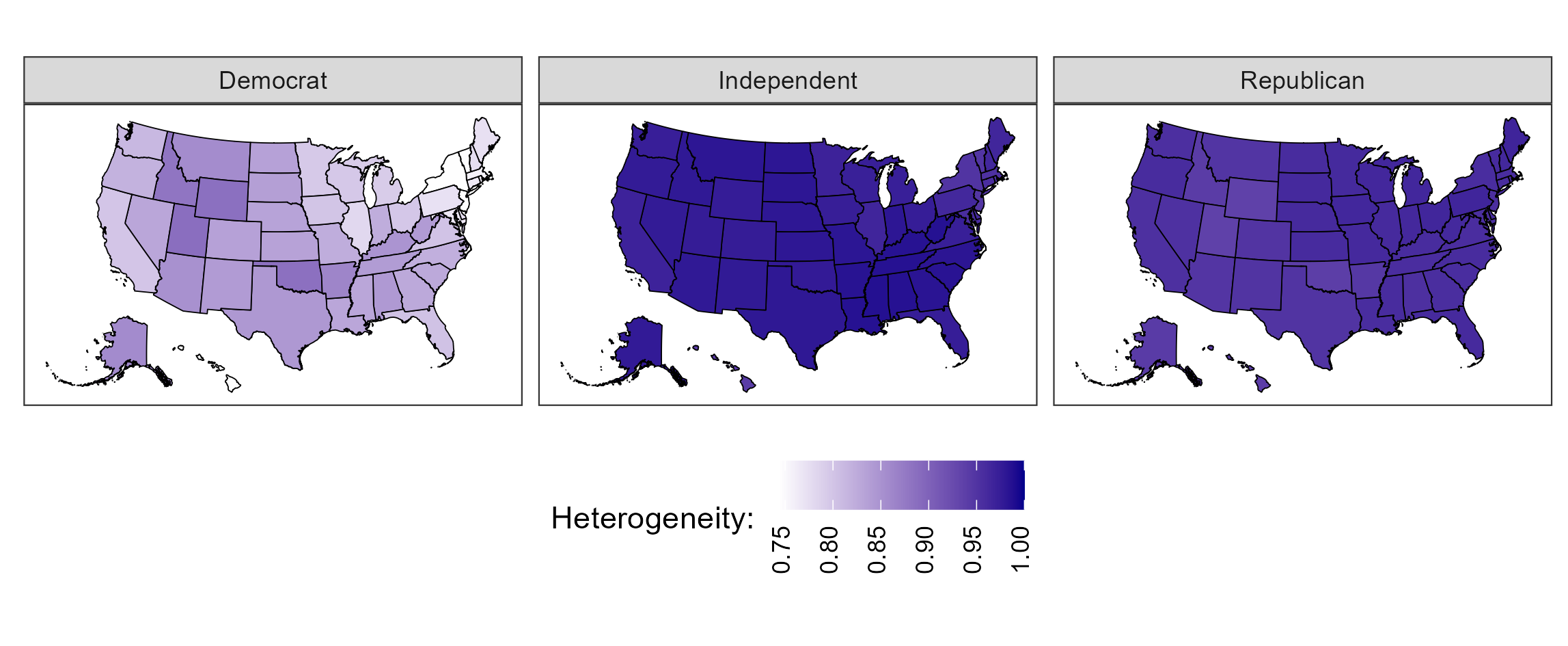}
	\label{fig:entropy}
	\caption*{\footnotesize \emph{Note:} The figure shows the entropy of state party groups. Higher values correspond to more heterogeneous views.}
\end{figure}

There are many quantities of interest that could be obtained using the joint distributions of opinion from \texttt{mvMRP}. Figure~\ref{fig:mapbyparty} shows one of the simplest, which is the joint distribution of support for both the Assault Weapons Ban and the Farm Bill across state-party groups. Note that although party is clearly an important predictor of opinion, there is also considerable geographic variation evident.

A second quantity one might be interested in is the heterogeneity of state-party groups. That is, within a state party, how variable is opinion across the four possible combinations of vote preferences. One way to measure this is to consider the entropy of the joint distribution of opinion in a state, i.e.
$H(\{p_i\}) = -\sum_i p_i \ln(p_i),$
which is minimized in a case where 100\% of a state-party supports the same combination of votes on the two bills and is minimized where 25\% of a state-party supports each combination. As such, higher values correspond to more heterogeneous views. Figure~\ref{fig:entropy} reports results by state-party, standardized to a range of 0 to 1. Here we can see that in general, Democrats hold more homogeneous views compared to Republicans and Independents. Again, there is some variation across states, with northeastern Democrats generally holding the most homogeneous views.

\subsection{Descriptive Statistics}
\label{app:responsiveness_descriptives}

Table~\ref{tab:descriptives} reports descriptive statistics for the variables included in the main empirical results. 

\begin{table}[!htb]
	\caption{Descriptive Statistics}
	\label{tab:descriptives}
	\begin{centering}
		
\begin{tabular}{@{\extracolsep{5pt}}lccccc} 
\\[-1.8ex]\hline 
\hline \\[-1.8ex] 
Statistic & \multicolumn{1}{c}{N} & \multicolumn{1}{c}{Mean} & \multicolumn{1}{c}{St. Dev.} & \multicolumn{1}{c}{Min} & \multicolumn{1}{c}{Max} \\ 
\hline \\[-1.8ex] 
Sen. Vote: Assault Ban & 96 & 0.39 & 0.49 & 0 & 1 \\ 
Sen. Vote: Farm Bill & 96 & 0.69 & 0.47 & 0 & 1 \\ 
Statewide Opinion & 96 & 28.41 & 9.45 & 11.88 & 46.87 \\ 
Independent Opinion & 96 & 29.05 & 6.77 & 15.17 & 40.68 \\ 
Copartisan Opinion & 96 & 33.35 & 14.80 & 8.51 & 56.43 \\ 
Party Share & 96 & 34.31 & 7.21 & 20.21 & 50.78 \\ 
President Share & 96 & 55.89 & 8.50 & 36.33 & 74.63 \\ 
\hline \\[-1.8ex] 
\multicolumn{6}{l}{} \\ 
\end{tabular} 

		\caption*{\footnotesize \emph{Note:} This table reports descriptive statistics for variables included in the empirical study.}
	\end{centering}
\end{table}

\subsection{Alternative Specification}
\label{app:sec_responsiveness_altspecs}

This section reports results for an alternative specification of the multidimensional responsiveness regression in the main text. Table~\ref{tab:altmodels} shows conditional logistic regression models for two specifications. The first column corresponds to the fifth model reported in Table~\ref{tab:empiricalregressions} of the main text, replicated here for comparison. The second column of Table~\ref{tab:altmodels} shows an alternative specification using the percent of each senator's state that identifies with their party, obtained from the \texttt{mvMRP} model. The interaction coefficients using the Party Share variable are similar to those using presidential vote, though only the interaction with copartisan opinion is statistically significant.

\begin{table}[!htb]
	\caption{Predicting Senators' Votes: Alternative Model}
	\label{tab:altmodels}
	\begin{centering}
		
\begin{tabular}{@{\extracolsep{5pt}}lcc} 
\\[-1.8ex]\hline \\[-1.8ex] 
\\[-1.8ex] & (1) & (2)\\ 
\hline \\[-1.8ex] 
 Independent Op. & 0.20$^{*}$ & 0.13 \\ 
  & (0.09) & (0.08) \\ 
  Independent Op. * Pres. Share & $-$0.01$^{**}$ &  \\ 
  & (0.004) &  \\ 
  Independent Op. * Party Share &  & $-$0.01 \\ 
  &  & (0.01) \\ 
  Copartisan Op. & 0.06$^{**}$ & 0.06$^{**}$ \\ 
  & (0.02) & (0.02) \\ 
  Copartisan Op. * Pres. Share & 0.01$^{**}$ &  \\ 
  & (0.003) &  \\ 
  Copartisan Op. * Party Share &  & 0.01$^{**}$ \\ 
  &  & (0.004) \\ 
 N & 96 & 96 \\ 
R$^{2}$ & 0.33 & 0.32 \\ 
Log Likelihood & $-$80.92 & $-$81.95 \\ 
\hline \\[-1.8ex] 
\multicolumn{3}{l}{} \\ 
\end{tabular} 

		\caption*{\footnotesize \emph{Note:} $^{*}$: $p < 0.05$; $^{**}$: $p < 0.01$. All models are conditional logistic regressions; the intercepts are not shown. All variables are measured on a scale of 0 to 100. For interpretability, party share is centered at 30\% and presidential share is centered at 50\%.}
	\end{centering}
\end{table}

Substantively, the models tell a similar story: As senators face greater electoral competition in the general election (i.e., in states where they have fewer copartisans or where their party's presidential candidate performed less well), they are more likely to respond to the preferences of independents. Where their parties are relatively strong, they are more responsive to the views of copartisans in the electorate.

\section{Software}\label{app:software}

This appendix provides additional information on the software used in this manuscript as well as a concise overview of other options.

\subsection{Software Demonstration}\label{app:software_demo}

We provide a brief demonstration of how to fit \texttt{mvMRP} using our software. The user must reshape their data to have one row for each observation $i$ and response category $\ell$. This can be easily done manually or using existing packages. After this, the Poisson regression is estimated with one line of code. Predictions can be done straightforwardly and turned into probabilities using a ``softmax.'' We provide some commentary in the accompanying code to clarify differences between \texttt{mvMRP} and MRP for researchers who have implemented MRP using, say, \texttt{lme4}. This model corresponds to the model with copartisanship in Section~\ref{sec:twostep_validation}. Adding income gives the model in Section~\ref{sec:application}.

\begin{lstlisting}
# Data has one row for each person (i) and response category (ell)
# response = 1 if the response category chosen, 0 otherwise

# v_fe(case_id) adds a FE for the respondent id

# Note that demvote and evang are *not* included as unregularized terms
# as they are not identified given the respondent FE
# Further, as choice is included as a FE, there is no need to include 
# a random intercept on policy, partyID or choice 
mvmrp.form <- response ~ v_fe(case_id) + choice + lag_copart +
(0 + demvote + evang + lag_copart | policy) + 
(0 + demvote + evang + lag_copart | partyID) + 
(0 + demvote + evang + lag_copart | choice) +
(1 | race : partyID) + (1 | race : policy) + (1 | race : choice) +
(1 | sex : partyID) + (1 | sex : policy) + (1 | sex : choice) +
(1 | educ : partyID) + (1 | educ : policy) + (1 | educ : choice) +
(1 | age : partyID) + (1 | age : policy) + (1 | age : choice) +
(1 | state : partyID) + (1 | state : policy) + (1 | state : choice) +
(1 | division : partyID) + (1 | division : policy) + 
(1 | division : choice)

# Estimating the mvMRP model
mvmrp.mdl <- vglmer(mvmrp.form, data = mvmrp.poll, family = 'poisson')

# Structure the post-stratification data
# is structured to have one row for each cell and response category. 
# Assume the variable "id" uniquely identifies each cell.

# Get the linear predictor for each cell
# This ignores the "v_fe" in the estim
poststrat$lp <- predict(mvmrp.mdl, data = poststrat)

# Using dplyr, turn the lp into probabilities using softmax
library(dplyr)
poststrat <- poststrat %>% group_by(id) %>%
mutate(prob = exp(lp - max(lp))/sum(exp(lp - max(lp))))
\end{lstlisting} 

\subsection{Existing Software Options}\label{app:software_other}

Existing software for multinomial or conditional logistic regression with random effects is rather limited. While \texttt{STAN} (\citealt{carpenter2017stan}) and especially implementations in \texttt{brms} (\citealt{burkner2017brms}) provides an easy-to-use interface---compatible with the \texttt{lme4} style formula used in the main text, they are a fully Bayesian approach and can be prohibitively expensive on large datasets and/or with many random effects (as we have in \texttt{mvMRP}). In terms of other non-fully Bayesian options, there are severe limitations to existing approaches. \texttt{lme4} (\citealt{bates2015lmer}) cannot estimate categorical outcomes; \texttt{mlogit} (\citealt{croissant2020mlogit}) cannot estimate hierarchical models except for mixed multinomial logistic regression (i.e., individual-specific random effects); \texttt{mgcv} (\citealt{wood2017gam}) cannot estimate conditional logistic regression, i.e. models with alternative-specific covariates. \texttt{mclogit} (\citealt{elff2025mclogit}) can estimate conditional logistic regressions with random effects, although it relies on techniques that are not guaranteed to converge (e.g., PQL) and are slow for similar reason to why traditional Laplace approximations are slower than their variational counterparts for large and complex models. Thus outside of its application to MRP, the methodology in this paper is a useful contribution to applied scholars. 

\subsection{Two-Stage MRP}\label{app:software_mgcv}

Our main comparison model in Section~\ref{sec:twostep_validation} is a two-stage MRP model. To estimate this, we use \texttt{mgcv} as it is the most flexible of the existing packages and is what, we think, an applied user who wished to do non-fully Bayesian estimation would employ.

We note that a limitation of \texttt{mgcv} is that the regularization---and thus the results---depend on the choice of baseline category. On the other hand, the model is more flexible than the implementation of our methodology discussed above as the random effect variance differs across category. The formula for the ``first stage'' model, predicting \texttt{pid3} is shown below, noting that in \texttt{mgcv}'s syntax, \verb|s(g, bs="re")| implements a random intercept. 
\begin{lstlisting}[language=R]
gam(pid3 ~ 
list(
s(race, bs = "re") + s(sex, bs = "re") + s(age, bs = "re") + 
s(educ, bs = "re") + s(state, bs = "re") + 
s(division, bs = "re") + demvote + evang,
s(race, bs = "re") + s(sex, bs = "re") + s(age, bs = "re") + 
s(educ, bs = "re") + s(state, bs = "re") + 
s(division, bs = "re") + demvote + evang)
), data = dat, family = multinom(K=2),
method = "REML"
)
\end{lstlisting}

The second stage is more straightforward and adds a random intercept to \texttt{partyID} to the model.

\begin{lstlisting}
gam(policy ~ s(pid3, bs = "re") + s(race, bs = "re") + 
s(sex, bs = "re") + s(age, bs = "re") + s(educ, bs = "re") + 
s(state, bs = "re") + s(division, bs = "re") + demvote + evang, 
data = dat_stage2, method = 'REML',
family = binomial())
\end{lstlisting}

The version with co-partisanship, explored in Appendix~\ref{app:validation_copart}, adds the corresponding lagged share of co-partisans to corresponding linear predictor as \texttt{mgcv} does not appear to allow for alternative-specific covariates. 

\end{document}